\RequirePackage{fix-cm}
\documentclass[twocolumn]{svjour3}          

\smartqed  

\usepackage{ucs}
\usepackage[utf8x]{inputenc}

\usepackage[numbers]{natbib}

\usepackage{graphicx}
\graphicspath{{images/}}

\usepackage{amsmath}
\usepackage{amsfonts}
\usepackage{relsize}
\usepackage{todonotes}
\usepackage{varwidth}
\usepackage[english]{babel}
\usepackage{subfig}                 
\usepackage{mathtools}
\usepackage{booktabs}
\usepackage{multirow}
\usepackage{tikz}
\usetikzlibrary{calc,arrows}

\newcommand{\tikzmark}[1]{%
  \tikz[overlay,remember picture] \node (#1) {};}

\usepackage[amssymb,binary]{SIunits}
\addunit{\flop}{Flop}
\usepackage{units}

\usepackage[plain]{algorithm}
\usepackage{algpseudocode}
\algrenewcommand{\algorithmiccomment}[1]{$//$ #1}
\algnewcommand\algorithmicto{\textbf{to}}
\algnewcommand\algorithmicforeach{\textbf{for\,each}}
\algblockdefx[Foreach]{Foreach}{EndForeach}%
	[1]{\algorithmicforeach\ #1\ \algorithmicdo}%
	{\algorithmicend\ \algorithmicforeach}

\usepackage{bm}
\let\vec\bm

\let\stdvec\vec
\newcommand{\vecdot}[1]{\dot{\stdvec{#1}}}
\newcommand{\mrcell}[2][c]{\begin{tabular}[#1]{@{}l@{}}#2\end{tabular}}
\newcommand{\argmin}{\operatornamewithlimits{arg\,min}}
\newcommand{\compl}{~\bot~}

\newcommand{\vectoscalar}[1]{{\renewcommand{\vec}{}#1}}

\newcommand{\vectovecdot}[1]{{\renewcommand{\vec}{\vecdot}#1}}

\newcommand{\card}[1]{\lvert#1\rvert}
\newcommand{\norm}[1]{\lVert#1\rVert}
\newcommand{\dt}{\ensuremath{\delta t}}
\newcommand{\overbar}[1]{\mkern 1.5mu\overline{\mkern-1.5mu#1\mkern-1.5mu}\mkern 1.5mu}
\newcommand{\bodyset}{\ensuremath{\mathcal{B}}}
\newcommand{\neighborset}{\ensuremath{\mathcal{N}}}
\newcommand{\contactset}{\ensuremath{\mathcal{C}}}
\newcommand{\procset}{\ensuremath{\mathcal{P}}}
\newcommand{\setprop}[2]{\ensuremath{\left\{ #1 \,\middle|\, #2 \right\}}}
\newcommand{\set}[1]{\ensuremath{\left\{ #1 \right\}}}

\newcommand{\intervalCC}[2]{\ensuremath{\left[ #1, #2 \right]}}

\newcommand{\mat}[1]{{\bf #1}}
\newcommand{\R}{\mathbb{R}}

\newcommand{\dvect}[2]{\ensuremath{\begin{pmatrix}#1\\#2\end{pmatrix}}}

\newcommand{\transp}{{\mathrm{T}}}
\newcommand{\cross}{\times}
\newcommand{\state}  [1]{\ensuremath{\vec{s}_{#1}}}
\newcommand{\pos}    [1]{\ensuremath{\vec{x}_{#1}}}
\newcommand{\orient} [1]{\ensuremath{\vec{\varphi}_{#1}}}
\newcommand{\linvel} [1]{\ensuremath{\vec{v}_{#1}}}
\newcommand{\angvel} [1]{\ensuremath{\vec{\omega}_{#1}}}
\newcommand{\postlinvel} [1]{\ensuremath{\vec{v}_{#1}^+}}
\newcommand{\postangvel} [1]{\ensuremath{\vec{\omega}_{#1}^+}}

\newcommand{\mass}   [1]{\ensuremath{m_{#1}}}
\newcommand{\inertia}[1]{\ensuremath{\mat{I}_{#1}}}

\newcommand{\posdot}   [1]{\ensuremath{\vectovecdot{\pos{#1}}}}
\newcommand{\orientdot}[1]{\ensuremath{\vectovecdot{\orient{#1}}}}

\newcommand{\force}    [1]{\ensuremath{\vec{f}_{#1}}}
\newcommand{\torque}   [1]{\ensuremath{\vec{\tau}_{#1}}}
\newcommand{\forceext} [1]{\ensuremath{\vec{f}_{#1,ext}}}
\newcommand{\torqueext}[1]{\ensuremath{\vec{\tau}_{#1,ext}}}

\newcommand{\cof}[1]{\ensuremath{\mu_{#1}}}

\newcommand{\contactforce}[1]{\ensuremath{\vec{\lambda}_{#1}}}
\newcommand{\contactforceapprox}[1]{\ensuremath{\vec{\tilde{\lambda}}_{#1}}}
\newcommand{\contactforceCFn}[1]{\ensuremath{\vectoscalar{\contactforce{#1,n}}}}
\newcommand{\contactforceCFto}[1]{\ensuremath{\contactforce{#1,to}}}

\newcommand{\discreterelvel}[1]{\ensuremath{{\vec{\delta v}_{#1}'}}}
\newcommand{\discreterelvelCFto}[1]{\ensuremath{\discreterelvel{#1,to}}}

\newcommand{\postrelvel}[1]{\ensuremath{{\vec{\delta v}_{#1}^+}}}
\newcommand{\postrelvelCFto}[1]{\ensuremath{\postrelvel{#1,to}}}
\newcommand{\postrelveldotCFto}[1]{\ensuremath{\vectovecdot{\postrelvelCFto{#1}}}}

\newcommand{\contactpos}    [1]{\ensuremath{\vec{\hat{x}}_{#1}}}
\newcommand{\contactposdot} [1]{\ensuremath{\vectovecdot{\contactpos{#1}}}}
\newcommand{\diag}{\operatornamewithlimits{diag}}
\newcommand{\numbodies}{\ensuremath{\nu_b}}
\newcommand{\numcontacts}{\ensuremath{\nu_c}}
\newcommand{\numsubdomains}{\ensuremath{\nu_p}}
\newcommand{\shape}    [1]{\ensuremath{\mathcal{S}_{#1}}}

\newcommand{\hull}    [1]{\ensuremath{\mathcal{H}_{#1}}}

\newcommand{\twodots}{\ensuremath{\mathrel{\ldotp\ldotp}}}

\DeclareMathAlphabet{\mathpzc}{OT1}{pzc}{m}{it}
\newcommand{\pe}{\ensuremath{\mathpzc{pe}}{}}

\newcommand{\shrinkeqnnew}[2]{\resizebox{#1\linewidth}{!}{\begin{varwidth}[t]{2\linewidth}\ensuremath{\displaystyle{#2}}\end{varwidth}}}

\newtheorem{requirement}{Requirement}

\journalname{Computational Particle Mechanics}

\begin{document}

\title{Ultrascale Simulations of Non-smooth Granular Dynamics}
\author{Tobias Preclik \and Ulrich R\"{u}de}


\institute{T. Preclik \at
           Lehrstuhl f\"{u}r Informatik 10 (Systemsimulation), Friedrich-Alexander-Universit\"{a}t Erlangen-N\"{u}rnberg, Cauerstr. 11, 91052 Erlangen, Germany \\
           \email{tobias.preclik@fau.de}
           \and
           U. R\"{u}de \at
           Lehrstuhl f\"{u}r Informatik 10 (Systemsimulation), Friedrich-Alexander-Universit\"{a}t Erlangen-N\"{u}rnberg, Cauerstr. 11, 91052 Erlangen, Germany \\
           \email{ulrich.ruede@fau.de}
}

\date{Received: 31.12.2014 / Accepted:}

\maketitle

\begin{abstract}
%
%
This article presents new 
algorithms 
for massively parallel granular dynamics simulations on
distributed memory architectures
using a domain partitioning approach.
Collisions are modelled with hard contacts
in order to hide their micro-dynamics and thus to extend
the 
time and length scales that can be simulated.
The multi-contact problem is solved using
a non-linear block Gauss-Seidel 
method that is conforming to 
the 
subdomain structure.
The parallel algorithms employ a sophisticated protocol between processors
that delegate algorithmic tasks such as contact treatment and position integration
uniquely and robustly to the processors.
%
Communication overhead is minimized through aggressive message aggregation,
leading to excellent strong and weak scaling.
The robustness
and 
scalability 
is assessed 
on three clusters including two peta-scale supercomputers 
with up to 458\,752 processor cores.
The simulations can reach unprecedented resolution of
up to ten billion ($10^{10}$) non-spherical particles and contacts.
\end{abstract}

\section{Introduction}
\label{sec:introduction}

Granular matter exhibits intriguing behaviours akin to solids, liquids or gases.
However, in contrast to those fundamental states of matter, granular matter
still cannot be described by a unified model equation homogenizing the
dynamics of the individual particles~\cite{mitarai12}. 
To date, 
the rich set of phenomena observed in granular matter,
can only be reproduced with simulations
that resolve every individual particle. 
In this paper, 
we will consider methods
where also the spatial
extent and geometric shape 
of the particles 
can be modelled.
Thus  
in addition to position and translational
velocity the
orientation and angular velocity  of each particle
constitute the state variables of the dynamical system.
The shapes of the
particles can be described for example by geometric primitives, such as spheres
or cylinders, with a low-dimensional parameterization.
Composite objects can be introduced 
as a set of primitives that are rigidly glued together.
Eventually, even meshes with a higher-dimensional
parameterization can be used.
In this article the shape of the particles does not change in time,
i.e.~no agglomeration,
fracture or deformation takes place.
The rates of change of the state variables
are 
described by the Newton-Euler equations, and
the particle interactions are determined by contact models.

Two
fundamentally different model types 
must be distinguished: Soft and hard contacts.
Soft contacts allow a local compliance in the contact region, whereas hard
contacts forbid penetrations. In the former class the contact forces can be
discontinuous in time, leading to non-differentiable but continuous velocities
after integration. The differential system can be 
cast e.g.\ as an ordinary
differential equation with a discontinuous right-hand side or as differential
inclusions. However, the resulting differential system is typically extremely stiff if
realistic material parameters are employed.

In the latter class, discontinuous
forces are not sufficient to accomplish non-penetration of the particles.
Instead, impulses are necessary to instantaneously change velocities on
collisions or in self-locking configurations if Coulomb friction is
present~\cite{shen11}. Stronger mathematical concepts are required to describe
the dynamics. For that purpose, Moreau introduced the measure differential
inclusions in~\cite{moreau88}. 

Hard contacts 
are an idealization of 
reality. The
rigidity of contacts has the advantage that the dynamics of the micro-collisions
does not have to be resolved in time. 
However, 
this also introduces ambiguities: The rigidity 
has the effect that the force chains along which a
particle is supported are no longer unique \cite{popa2014regularized}. If energy is dissipated, this
also effects the dynamics. To 
integrate measure differential inclusions
numerically
in time, two options exist: In the first approach 
the integration is performed in 
subintervals from one impulsive event to the next~\cite{miller04,esefeld14}.
At each event an instantaneous impact problem 
must be solved whose solution
serves as initial condition of the subsequent integration 
subinterval. Impact problems
can range from simple binary collisions, to self-locking configurations, to
complicated instantaneous frictional multi-contact problems with simultaneous
impacts. The dynamics between events are described by differential inclusions,
differential algebraic equations or ordinary differential equations.
Predicting the times of the upcoming events correctly is non-trivial
in general and handling them in order in parallel is impeding the scalability~\cite{miller04}.
In the second approach no efforts are made to detect events, but the contact conditions are
only required to be satisfied at discrete points in time. This approach is
commonly referred to as a time-stepping method.

This article focuses on the treatment of hard contacts in order to avoid
the temporal resolution of micro-collisions and thus the dependence of the
time-step length on the stiff\-ness of the contacts. In order to avoid
the resolution of events a time-stepping method is employed.
This considerably pushes the time scales accessible to granular flow
simulations for stiff contacts.

To estimate the order of a typical real-life problem size of a granular system,
consider an excavator bucket
with a capacity of $\unit[1]{\meter\cubed}$. Assuming sand grains with a diameter of $\unit[0.15]{\milli\meter}$,
and assuming that they 
are packed with a solid volume fraction of 0.6, the
excavator bucket contains in the order of $10^{10}$ particles. In such a
dense packing the number of contacts is in the same order as the number of
particles. 
Only large scale parallel systems with distributed memory can
provide enough memory to store
the data 
and provide sufficient computational power to integrate such systems
for a relevant simulation time. 
Consequently a massive parallelization of the numerical method for architectures
with distributed memory is absolutely essential.

In the last half decade several approaches were published suggesting
parallelizations of the methods integrating the equations of motion of
rigid particles in hard contact \cite{visseq12,visseq13,koziara11,shojaaee12,iglberger2009massively,iglberger2010massively,negrut2012leveraging}.
The approach put forward in this article builds conceptually 
on these previous approaches but exceeds them substantially
by consistently parallelizing all parts of the code,
consistently distributing all simulation data (including the description of the
domain partitioning), systematically minimizing the 
volume of communication 
and the number
of exchanged messages, and relying exclusively on efficient nearest-neighbor
communication. 
The approach described here additionally spares 
the expensive assembly of system matrices by
employing matrix-free computations. 
All this is accomplished without sacrificing accuracy.
The matrix-freeness allows the direct and straight forward evaluation of
wrenches in parallel and thus reduces the amount of communicated data.
Furthermore, an exceptionally robust synchronization protocol is defined,
which is not susceptible to numerical errors. The excellent parallel scaling
behaviour is then demonstrated for dilute and dense test problems in strong-
and weak-scaling experiments on three clusters with fundamentally different
interconnect networks. Among the test machines are the peta-scale supercomputers
SuperMUC and Juqueen. The results show that given a sufficient computational
intensity of the granular setup and an adequate interconnect, few hundred
particles per process are enough to obtain 
satisfactory scaling even on millions of processes.

In Sect.~\ref{sec:continuous_dynamical_system} of this paper the underlying differential
equations and the time-continuous formulation of the hard contact models
are formulated. Sect.~\ref{sec:discrete_dynamical_system} proposes a
discretization scheme and discrete constraints for the hard contact model.
The problem of reducing the number of contacts in the system for efficiency
reasons is addressed in Sect.~\ref{sec:contact_detection}. Subsequently,
an improved numerical method for solving multi-contact problems in parallel
is introduced in Sect.~\ref{sec:numerical_solution_algorithms} before turning to the
design of the parallelization in Sect.~\ref{sec:parallelization_design}. 
The
scalability of the parallelization is then demonstrated in Sect.~\ref{sec:scaling_experiments}
by means of dilute and dense setups on three different clusters. Finally, the algorithms and results are
compared to previous work by other authors in Sect.~\ref{sec:related_work} before summarizing in Sect.~\ref{sec:summary}.

\section{Continuous Dynamical System}
\label{sec:continuous_dynamical_system}

The Newton-Euler equations for a system with $\numbodies$~particles are~\cite{liu12}
\begin{equation*}
	\shrinkeqnnew{1}{
		\begin{split}
			\dvect{\posdot{}(t)}{\orientdot{}(t)} & =
			\dvect{\linvel{}(t)}{\mat{Q}(\orient{}(t)) \angvel{}(t)}, \\
			\mat{M}(\orient{}(t)) \dvect{\dot{\linvel{}}(t)}{\dot{\angvel{}}(t)} & =
			\dvect{ \force{}(\state{}(t), t)}{\torque{}(\state{}(t), t) - \angvel{}(t) \times \inertia{}(\orient{}(t)) \angvel{}(t)},
		\end{split}
	}
\end{equation*}
where the positions~$\pos{}(t) \in \R^{3\numbodies}$, the rotations~$\orient{}(t) \in \R^{4\numbodies}$, translational
velocities~$\linvel{}(t) \in \R^{3\numbodies}$, and angular velocities $\angvel{}(t) \in \R^{3\numbodies}$ are the state
variables at time~$t$.

Different parameterizations exist for the rotations, but quaternions
having four real components are the parameterization of choice here. Independent of the
parameterization, the derivatives of the rotation components can be expressed in terms of a matrix-vector
product between a block-diagonal matrix and the angular velocities~\cite{diebel06}.
If the rotation of particle~$i$ is described by the quaternion $q_w + q_x \mathrm{i} + q_y \mathrm{j} + q_z \mathrm{k} \in \mathbb{H}$ then, according to~\cite{diebel06}, the $i$-th diagonal block of $\mat{Q}(\orient{}(t))$ is
\begin{equation*}
	\mat{Q}_{ii}(\orient{i}(t)) = \frac{1}{2} \begin{bmatrix*}[r] -q_x & -q_y & -q_z \\ q_w & q_z & -q_y \\ -q_z & q_w & q_x \\ q_y & -q_x & q_w \end{bmatrix*}.
\end{equation*}

Each particle has an associated body frame
whose origin coincides with the body's center of mass and whose axes are
initially aligned with the axes of the observational frame. The body frame is
rigidly attached to the body and translates and rotates with it. All of the
state variables and other quantities are expressed in the observational frame unless noted otherwise.
Furthermore, the matrix
\begin{equation*}
	\mat{M}(\orient{}(t)) = \begin{bmatrix} \diag\limits_{i = 1 \twodots \numbodies} \mass{i}\mat{1} & \\ & \diag\limits_{i = 1 \twodots \numbodies} \inertia{ii}(\orient{i}(t)) \end{bmatrix}
\end{equation*}
is the block-diagonal mass matrix, where $\mat{1}$ denotes the $3 \times 3$
identity matrix. The mass matrix contains the constant particle masses $\mass{i}$ and
the particles' inertia matrices $\inertia{ii}(\orient{i}(t))$ about the particles' centers of mass.
The latter can be calculated by similarity transformations from the constant body frame
inertia matrices $\inertia{ii}^0$. If the body frames are attached such that they coincide with
the principal axes of their particles, then the body frame inertia matrices are
diagonal, and floating-point operations as well as memory can be saved. The lower-right
block of the mass matrix corresponds to the matrix $\inertia{}(\orient{}(t))$.
$\force{}(\state{}(t), t)$ and $\torque{}(\state{}(t), t)$ are the total forces
and torques (together they are referred to as wrenches) acting at the particles'
centers of mass. Both may depend on any of the state variables~$\state{}(t)$ of
the system and time~$t$. The wrench contributions from contact reactions are
summed up with external forces $\vec{f}_{ext}$ and torques $\vec{\tau}_{ext}$ such as fictitious forces from
non-inertial reference frames.

Let $\contactforce{j}(t) \in \R^3$ be the
contact reaction of a contact~$j \in \contactset$, where $\contactset = \set{1 \twodots \numcontacts}$
is the set of potential contact indices. Let $(j_1, j_2) \in \bodyset^2$ be the
index pair of both particles involved in the contact~$j$, where $\bodyset = \set{1 \twodots \numbodies}$
is the set of body indices. Let $\contactpos{j}(\pos{}(t), \orient{}(t)) \in \R^3$
be the location of contact~$j$, then the wrench on body~$i$ is
%
\begin{equation}
	\label{eq:wrench}%
	\shrinkeqnnew{0.9}{
		\begin{split}
			\begin{pmatrix} \force{i}(\state{}(t), t) \\ \torque{i}(\state{}(t), t) \end{pmatrix} = \begin{pmatrix} \forceext{i}(\state{}(t), t) \\ \torqueext{i}(\state{}(t), t) \end{pmatrix}
			& + \sum_{\substack{j \in \contactset\\j_1 = i}} \begin{bmatrix} \mat{1} \\ (\contactpos{j}(\pos{}(t), \orient{}(t)) - \pos{i}(t))^\cross \end{bmatrix} \contactforce{j}(t) \\
			& \underbrace{- \sum_{\substack{j \in \contactset\\j_2 = i}} \begin{bmatrix} \mat{1} \\ (\contactpos{j}(\pos{}(t), \orient{}(t)) - \pos{i}(t))^\cross \end{bmatrix} \contactforce{j}(t)}_{\text{wrench contributions}},
		\end{split}
	}
\end{equation}
where $\left(\cdot\right)^\cross$ is a
matrix, which when multiplied to a vector corresponds to the
cross product between its operand $\left(\cdot\right)$ and the vector.

In contrast to soft contact models,
the contact reactions in hard contact
models cannot be explicitly expressed as a function of the state variables but
are defined implicitly, e.g.\ by implicit non-linear functions~\cite{leyffer06}, complementarity
constraints~\cite{anitescu97,anitescu10}, or inclusions~\cite{studer09}. In any
case, the constraints distinguish between reactions in the directions normal to the contact surfaces
and reactions in the tangential planes of the contact surfaces. The former are used
to formulate the non-penetration constraints, and the latter are used to
formulate the friction constraints. For that reason, each contact~$j$ is associated
with a contact frame, where the axis $\vec n_j(\pos{}(t), \orient{}(t)) \in \R^3$ points along the
direction normal to the contact surface, and the other two axes
$\vec t_j(\pos{}(t), \orient{}(t)) \in \R^3$ and $\vec o_j(\pos{}(t), \orient{}(t)) \in \R^3$ span the tangential plane of
the contact.

Let $\shape{i}$ be the set of points in the observational frame
defining the shape of particle~$i$, and let $f_i(\pos{i}(t), \orient{i}(t), \vec y) \in \R$ be the
associated signed distance function for a point $\vec y$ in the observational frame.
The signed distance function shall be negative in the interior of $\shape{i}$.
Assuming
that all particles are (strictly) convex with sufficiently smooth boundaries,
then for a pair of particles $(j_1, j_2)$ involved in a contact~$j$, the contact
location $\contactpos{j}(\pos{}(t), \orient{}(t))$ is defined by the optimization problem
\begin{equation}
	\label{eq:contactpos}
	\begin{split}
		\contactpos{j}(t) & := \contactpos{j}(\pos{}(t), \orient{}(t)) \\
		& = \argmin_{f_{j_2}(\pos{j_2}(t), \orient{j_2}(t), \vec y) \leq 0} f_{j_1}(\pos{j_1}(t), \orient{j_1}(t), \vec y),
	\end{split}
\end{equation}
with associated contact normal
\begin{equation*}
	\shrinkeqnnew{1}{\vec n_j(t) := \vec n_j(\pos{}(t), \orient{}(t)) = \nabla_{\vec y} f_{j_2}(\pos{j_2}(t), \orient{j_2}(t), \contactpos{j}(t)),}
\end{equation*}
pointing outwards with respect to $\shape{j_2}$ and associated signed contact
distance
\begin{equation*}
	\shrinkeqnnew{1}{\xi_j(t) := \xi_j(\pos{}(t), \orient{}(t)) = f_{j_1}(\pos{j_1}(t), \orient{j_1}(t), \contactpos{j}(t))}
\end{equation*}
which is negative 
in the case of penetrations.

For convex particles each pair of bodies results in a potential
contact, and thus the total number of contacts $\numcontacts$ is limited by
$\frac{1}{2}\numbodies(\numbodies - 1)$. Non-convex objects e.g.\ can be
implemented as composite objects of convex particles. By convention a
positive reaction in normal direction is repulsive, and thus the
contact reaction $\contactforce{j}(t)$ acts positively on particle $j_1$
and negatively on $j_2$, thus explaining the signs in~\eqref{eq:wrench}.
By applying the opposite reactions at the same point in the observational
frame, not only the linear momentum can be conserved but also the angular
momentum of the system. Conservation of energy can only hold if the contact
model does not include dissipative 
effects.
Hard-contact models require the Signorini condition to hold. Written as
a complementarity condition for a contact~$j$, it reads
\begin{equation*}
	\xi_j(t) \geq 0 \compl \contactforceCFn{j}(t) \geq 0,
\end{equation*}
where $\contactforceCFn{j}(t) = \vec n_j(t)^\transp \contactforce{j}(t)$.
The signed contact distance is required to be non-negative, resulting in a
non-penetration constraint. The contact reaction in direction of the contact
normal is also required to be non-negative, resulting in non-adhesive contact
reactions. Furthermore, both quantities must be complementary, meaning that
either of them must be equal to zero. This effects that the contact reaction can
only be non-zero if the contact is closed.

However, the Signorini condition does
not determine the contact reaction force if the contact is closed. In that
case the non-penetration constraint on the velocity level,
\begin{equation*}
	\dot{\xi}_j^+(t) \geq 0 \compl \contactforceCFn{j}(t) \geq 0,
\end{equation*}
%
must be added to the system, where $\dot{\xi}_j^+$ is the right derivative
of the signed contact distance with respect to time. The constraint allows the
contact to break only if no reaction force is present and otherwise forces
$\dot{\xi}_j^+(t) = 0$. In the latter case the reaction force is still not
fixed. The non-penetration constraint on the acceleration level,
\begin{equation*}
	\ddot{\xi}_j^+(t) \geq 0 \compl \contactforceCFn{j}(t) \geq 0,
\end{equation*}
then determines the force also if $\ddot{\xi}_j^+(t) = 0$. When
considering impacts, a non-penetration constraint for the reaction impulse in
the direction normal to the contact surface must be formulated, and, if the
contact is closed, an additional constraint modelling an impact law such as
Newton's impact 
must be added.

These non-penetration conditions can be complemented by a friction
condition. The most prominent model for dry frictional contact is the Coulomb
model which restricts the relative contact velocity in the tangential plane of
the contact. The relative contact velocity for a pair of particles $(j_1, j_2)$
involved in a contact~$j$ is
\begin{equation*}
	\begin{split}
		\postrelvel{j}(\state{}(t)) = & \postlinvel{j_1}(t) + \postangvel{j_1}(t) \cross (\contactpos{j}(\pos{}(t), \orient{}(t)) - \pos{j_1}(t)) \\
								- & \postlinvel{j_2}(t) - \postangvel{j_2}(t) \cross (\contactpos{j}(\pos{}(t), \orient{}(t)) - \pos{j_2}(t)).
	\end{split}
\end{equation*}
Let
\begin{equation*}
	\postrelvelCFto{j}(t) := \postrelvelCFto{j}(\state{}(t)) = \dvect{\vec t_j(\pos{}(t), \orient{}(t))^\transp \postrelvel{j}(\state{}(t))}{\vec o_j(\pos{}(t), \orient{}(t))^\transp \postrelvel{j}(\state{}(t))}
\end{equation*}
be the relative contact velocity in the tangential plane after application of the contact impulses, then the Coulomb
conditions for a non-impulsive point in time~$t$ are
\begin{equation*}
	\begin{split}
		& \norm{\contactforceCFto{j}(t)}_2 \leq \cof{j} \contactforceCFn{j}(t)\ \text{and} \\
		& \norm{\postrelvelCFto{j}(t)}_2 \contactforceCFto{j}(t) = - \cof{j} \contactforceCFn{j}(t) \postrelvelCFto{j}(t).
	\end{split}
\end{equation*}
However, if $\norm{\postrelvelCFto{j}(t)}_2 = 0$ these conditions 
must be
supplemented by the constraint
\begin{equation*}
	\norm{\postrelveldotCFto{j}(t)}_2 \contactforceCFto{j}(t) = - \cof{j} \contactforceCFn{j}(t) \postrelveldotCFto{j}(t)
\end{equation*}
on acceleration level in order to determine the friction force. Likewise constraints
for the friction impulse are necessary. At this point we refrain from
formulating the measure differential inclusion in detail since it would not
contribute information essential to the remaining paper which only deals
with the discrete-time system.

\section{Discrete Dynamical System}
\label{sec:discrete_dynamical_system}

In simulations of granular matter impulsive reactions are abundant. Higher-order
integrators for time-stepping schemes are still subject to active
research~\cite{schindler14}. In particular, discontinuities
pose problems for these integrators. Hence, the continuous dynamical system
is discretized in the following with an integrator of order one, resembling the
semi-implicit Euler method and similar to the one suggested in~\cite{anitescu02}.

Let $\state{}$, $\pos{}$, $\orient{}$, $\linvel{}$ and $\angvel{}$ denote the
given discrete-time
state variables at time~$t$ and $\contactforce{}$ the contact reactions at
time~$t$. Then the state variables at time~$t+\dt$ are functions depending
on the contact reactions: $\state{}'(\contactforce{})$, $\pos{}'(\contactforce{})$, $\orient{}'(\contactforce{})$, $\linvel{}'(\contactforce{})$ and
$\angvel{}'(\contactforce{})$. The discrete-time Newton-Euler
equations integrated by the proposed scheme are
\begin{equation}
	\label{eq:integrationscheme}
	\shrinkeqnnew{0.9}{
		\begin{split}
			\dvect{\pos{}'(\contactforce{})}{\orient{}'(\contactforce{})}    & = \dvect{\pos{}}{\orient{}}    + \dt \dvect{\linvel{}'(\contactforce{})}{\mat{Q}(\orient{}) \angvel{}'(\contactforce{})}, \\
			\dvect{\linvel{}'(\contactforce{})}{\angvel{}'(\contactforce{})} & = \dvect{\linvel{}}{\angvel{}} + \dt \mat{M}(\orient{})^{-1} \dvect{ \force{}(\state{}, \contactforce{}, t)}{\torque{}(\state{}, \contactforce{}, t) - \angvel{} \times \inertia{}(\orient{}) \angvel{}}.
		\end{split}
	}
\end{equation}
Positions and orientations at time $t + \dt$ appear exclusively on the
left-hand side of the position and orientation integration. Velocities at time
$t + \dt$ appear on the left-hand side of the velocity integration and
additionally in the integration of positions and orientations. The numerical
integration of the quaternion has the effect that the quaternion gradually looses
its unit length. This deficiency can be compensated by renormalizing the
quaternions after each integration.

Instead of discretizing each of the five intermittently active continuous-time
complementarity constraints, the Signorini condition is only
required to hold at the end of each time step. This has the effect that impulsive reactions
are no longer necessary to satisfy the condition since the condition is
no longer required to be fulfilled instantaneously. Furthermore, the signed
distance function gets linearized, resulting in
\begin{equation*}
	\xi_j(t + \dt) = \xi_j(t) + \dt \dot{\xi}_j(t) + \mathcal{O}(\dt^2),
\end{equation*}
where the time derivative of the signed contact distance can be determined
to be
%
\begin{equation*}
	\dot{\xi}_j(t) = \vec n_j(t)^\transp \postrelvel{j}(\state{}(t))
\end{equation*}
under the assumption that the contact point $\contactpos{j}(t)$ translates and
rotates in accordance with body $j_2$, such that
\begin{equation*}
	\contactposdot{j}(t) = \linvel{j_2}(t) + \angvel{j_2}(t) \cross (\contactpos{j}(t) - \pos{j_2}(t)).
\end{equation*}
Let the time-discrete relative contact velocity be
\begin{equation*}
	\begin{split}
		\discreterelvel{j}(\contactforce{}) = & \linvel{j_1}'(\contactforce{}) + \angvel{j_1}'(\contactforce{}) \cross (\contactpos{j} - \pos{j_1}) \\
		                                    - & \linvel{j_2}'(\contactforce{}) - \angvel{j_2}'(\contactforce{}) \cross (\contactpos{j} - \pos{j_2}),
	\end{split}
\end{equation*}
where the velocities are discretized implicitly. The discrete non-penetration constraint then is
\begin{equation}
	\label{eq:nonpenetration}
	\frac{\xi_j}{\dt} + \vec n_j^\transp \discreterelvel{j}(\contactforce{}) \geq 0 \compl \contactforceCFn{j} \geq 0.
\end{equation}
The term $\frac{\xi_j}{\dt}$ acts as an error correction term if penetrations
are present ($\xi_j < 0$). In that case it can be scaled down to avoid introducing an
excessive amount of energy. If no numerical error is present, the contact
is inelastic. The frictional constraints translate into
\begin{equation}
	\label{eq:coulomb}
	\begin{split}
		& \norm{\contactforceCFto{j}}_2 \leq \cof{j} \contactforceCFn{j}\ \text{and} \\
		& \norm{\discreterelvelCFto{j}(\contactforce{})}_2 \contactforceCFto{j} = - \cof{j} \contactforceCFn{j} \discreterelvelCFto{j}(\contactforce{}).
	\end{split}
\end{equation}

Let $\vec F_j(\contactforce{}) = \vec 0$ denote a non-linear system of equations
equivalent to the constraints from \eqref{eq:nonpenetration} and \eqref{eq:coulomb}
of a single contact~$j$, and let $\vec F(\contactforce{})$ denote the collection
of all $\vec F_j(\contactforce{})$. Neither $\vec F(\contactforce{}) = \vec 0$
nor $\vec F_j(\contactforce{}) = \vec 0$ for given $\contactforce{\overbar{j}}$ have
unique solutions. Let $\vec F_j^{-1}(\vec 0, \contactforce{\overbar{j}})$
be a possible solution of the one-contact problem of contact~$j$, given
the contact reactions $\contactforce{\overbar{j}}$ of all other contacts~$\overbar{j}$.

A detailed discussion of solution algorithms for one-contact problems is out of
the scope of this article. However, splitting methods, where non-penetration
and friction constraints are solved separately, are prone to
slow convergence or cycling. In~\cite{bonnefon11} Bonnefon et al.\ solve the
one-contact problem by finding the root of a quartic polynomial.
Numerous other approaches exist for modified friction laws, notably those
where the friction cone is approximated by a polyhedral cone and solution
algorithms for linear complementarity problems can be used~\cite{anitescu97,sauer98}. In
any case the algorithm of choice should be extremely robust in order to
successfully resolve $\numcontacts$ contacts per iteration and time step, where
$\numcontacts$ can be in the order of $10^{10}$ in this article.

\section{Contact Detection}
\label{sec:contact_detection}

The contact problem $\vec F(\contactforce{}) = \vec 0$ has
$\mathcal{O}(\numbodies^2)$ non-linear equations. Thus, already the
setup of the contact problem would not run in linear time, much less
the solution algorithm even if it 
were optimal. The contact
constraints of a contact~$j$ can be removed from the system without
altering the result if the contact is known to stay
open~($\contactforce{j} = \vec 0$) within the current time step.
Let
\begin{equation*}
	\shape{i}(t) = \setprop{\vec y \in \R^3}{f_i(\pos{i}(t), \orient{i}(t), \vec y) \leq 0}
\end{equation*}
be the set of points in space corresponding to the rotated and translated shape of particle~$i$ at time~$t$
and let
\begin{equation*}
	\hull{i}(t) = \shape{i}(t) + \setprop{\vec y \in \R^3}{\norm{\vec y}_2 \leq h_i(t)}
\end{equation*}
be an intersection hull that spherically expands the particle shape by the
radius~$h_i(t) > 0$. If $h_i(t)$ is chosen large enough then an algorithm
finding intersections between the hulls can detect all contacts that
can potentially become active in the current time step. A possible choice
for the expansion radius is
\begin{equation}
	\label{eq:expansionradius}
	h_i(t) = \dt ( \norm{\linvel{i}(t)}_2 + \norm{\angvel{i}(t)}_2 \overline{r}_i ) + \tau,
\end{equation}
where $\overline{r}_i = \max_{\vec y \in \shape{i}(0)} \norm{\vec y}_2$ is the
bounding radius of particle~$i$, and $\tau$ is a safety margin. The safety
margin becomes necessary since an explicit Euler step is underlying the
derivation of \eqref{eq:expansionradius}. In practice,
the usage of intersection hulls reduces the number of contacts considerably. E.g.,\ monodisperse spherical
particles can have at most 12~contacts per particle if the expansion radii are small enough~\cite{schuette52}, resulting in $\mathcal{O}(\numbodies)$
potential contacts.

Broad-phase contact detection algorithms
aim to find as few as possible candidate particle pairs for contacts by using
e.g.\ spatial partitioning approaches or exploiting temporal coherence of the
particle positions~\cite{cohen95}. The candidate pairs are then checked in detail in the
narrow-phase contact detection, where \eqref{eq:contactpos} is solved for
each pair, leading to the contact location $\contactpos{j}$, normal
$\vec n_j$ and signed distance $\xi_j$ for a contact~$j$. 

To solve
\eqref{eq:contactpos} for non-overlapping particles, the
Gil\-bert-John\-son--Keer\-thi~(GJK) algorithm can be used~\cite{gilbert88,bergen99}.
For
overlapping particle shapes the expanding polytope algorithm~(EPA) computes
approximate solutions~\cite{bergen01}. 
For simple geometric primitives like
spheres, the optimization problem can be solved analytically. The indices of all
contacts found that way form the set of potential contacts
$\contactset = \set{1 \twodots \numcontacts}$ at time~$t$.
Let $\vec{F}(\contactforce{}) = \vec 0$ from now on denote the
contact problem where all contact conditions and contact reactions whose
indices are not part of $\contactset$ have been filtered out.

\section{Numerical Solution Algorithms}
\label{sec:numerical_solution_algorithms}

To solve the multi-contact problem, when suitable solution algorithms for the
one-contact problems $\vec F_j^{-1}$ are given, a non-linear block Gauss-Seidel~(NBGS)
can be used as propagated by the non-smooth contact dynamics~(NSCD)
method~\cite{jean99}. Unfortunately, the Gauss-Seidel algorithm cannot
be efficiently 
executed in parallel for 
irregular data dependencies as
they appear in contact problems~\cite{koziara11}. 

As an alternative, a more general variant is
proposed here, accommodating the subdomain structure that will arise in the
domain partitioning.
Therefore, each contact~$j \in \contactset$ is associated with
a subdomain number $s_c(j) \in \procset$,
where $\procset = \set{1 \twodots \numsubdomains}$ is the set of subdomain indices
for $\numsubdomains$~subdomains. Alg.~\ref{alg:sdnbgs} presents pseudo-code
for the subdomain NBGS with the relaxation parameter~$\omega > 0$.
\begin{algorithm}
	\begin{algorithmic}[1]
		\State $k \gets 0$
		\State $\contactforce{}^{(k)} \gets \vec 0$
		\While{convergence criterion not met} \label{line:loopheader}
			\For{$j \gets 1\ \algorithmicto\ \numcontacts$}
				\For{$l \in \contactset$}
					\State $\contactforceapprox{l}^{(k,j)} \gets \begin{cases} \contactforce{l}^{(k+1)} & \text{if $l < j \land s_c(l) = s_c(j)$} \\ \contactforce{l}^{(k)} & \text{else} \end{cases}$
				\EndFor
				\State $\vec y \gets \vec F_j^{-1}(\vec 0, \contactforceapprox{\overbar{j}}^{(k,j)})$ \label{line:solve}
				\State $\contactforce{j}^{(k+1)} \gets \omega \vec y + (1 - \omega) \contactforce{j}^{(k)}$ \label{line:update}
			\EndFor
			\State $k \gets k + 1$
		\EndWhile
	\end{algorithmic}
	\caption{The subdomain~NBGS method with relaxation parameter $\omega$.\label{alg:sdnbgs}}
\end{algorithm}
The initial solution is chosen to be zero, however, any other initialization
can be used, in particular contact reactions from the previous time step.

The algorithm is of iterative nature and needs an appropriate stopping
criterion to terminate. In each iteration~$k$ a sweep over all contacts is
performed, where each contact~$j$ is relaxed, given an approximation of all other
contact reactions $\contactforceapprox{}^{(k,j)}$.
In the subdomain NBGS, the approximation of contact
reaction~$l$ is taken from the current iteration if it was already
relaxed~($l < j$) and if it is associated with the same subdomain as
the contact~$j$ to be relaxed~($s_c(l) = s_c(j)$). In all other cases, the
approximation is taken from the previous iteration. The contact reaction
$\contactforce{j}^{(k+1)}$ is then a weighted mean between the previous
approximation and the relaxation result. If all contacts are associated
with the same subdomain and $\omega = 1$ then Alg.~\ref{alg:sdnbgs}
corresponds to a classic NBGS. If each contact is associated to a different
subdomain then Alg.~\ref{alg:sdnbgs} corresponds to a non-linear block
Jacobi~(NBJ) with relaxation parameter~$\omega$.

\section{Parallelization Design}
\label{sec:parallelization_design}

Sect.~\ref{sec:domain_partitioning} introduces the domain partitioning
approach. Sect.~\ref{sec:shadow_copies} then discusses requirements that
must be met in order to be able to treat all contacts exactly once in
parallel. Sect.~\ref{sec:accumulator_and_correction_variables} explains
how accumulator and correction variables can be used in order to reduce
data dependencies to other processes. In Sect.~\ref{sec:nearest_neighbor_communication}
conditions are discussed under which the set of communication partners
can be reduced to the nearest neighbors. Time-integration and the subsequent
necessity of synchronization are addressed in Sect.~\ref{sec:time-integration_and_synchronization}
before summarizing the time-stepping procedure in Sect.~\ref{sec:parallelization_summary}.

\subsection{Domain Partitioning}
\label{sec:domain_partitioning}

Under the assumption that no contacts are present, there exists no coupling
between the data of any two particles, and the problem becomes embarrassingly
parallel: Each process integrates $\lfloor \frac{\numbodies}{\numsubdomains} \rfloor$
or $\lceil \frac{\numbodies}{\numsubdomains} \rceil$ particles.
Let $s_b(i) \in \procset$
determine the process responsible for the time-integration of particle~$i$ as of now
referred to as the parent process. All data associated with this particle, that is the
state variables (position, orientation, velocities) and constants (mass, body
frame inertia matrix, shape parameters), are instantiated only at the parent
process in order to distribute the total memory load. However, contacts or
short-range potentials introduce data dependencies
to particles that in general are not instantiated on the local process nor on a process
close to the local one, rendering a proper scaling impossible. A domain
partitioning approach alleviates this problem.

Let $\Omega$ denote the computational domain within which all particles are
located and $\Omega_p \subseteq \Omega,  p \in \procset$,
a family of disjoint subdomains
into which the domain shall be partitioned. In this connection subdomain boundaries are
associated to exactly one process.
One process shall be executed per subdomain. The
number of processes can e.g.\ correspond to the number of compute nodes in a
hybrid parallelization or to the total number of cores or even threads in a
homogeneous parallelization. In the domain partitioning approach the
integration of a particle whose center of mass $\pos{i}$ is located in a
subdomain $\Omega_p$ at time~$t$ is calculated by process~$p$. That way
data dependencies typically pertain the local or neighboring subdomains since
they are considered to be of short range. Let $s_b(i)$ be adapted accordingly.
Special care is required when associating a particle to a subdomain whose center
of mass is located on or near subdomain interfaces. Especially, periodic
boundary conditions can complicate the association process since the finite
precision of floating-point arithmetics does in general not allow a consistent parametric
description of subdomains across periodic boundaries. Sect.~\ref{sec:time-integration_and_synchronization} explains
how the synchronization protocol can be used to allow a reliable association.

The domain partitioning should be chosen such that an 
equal number of particles
is located initially in each subdomain and sustained over the course of the
simulation in order to balance the computational load which is directly
proportional to the number of particles. Particles now migrate between processes
if their positions change the subdomain. Migration can lead to severe load imbalances
that may need to be addressed by dynamically repartitioning the domain. Such
load-balancing techniques are beyond the scope of this article.

\subsection{Shadow Copies}
\label{sec:shadow_copies}

A pure local instantiation of particles has the effect 
that contacts cannot be detected  between particles
that are not located on the same process. 
A process can
detect a contact
if both particles involved in the contact are instantiated on
that process. In order to guarantee that at least one process can detect a
contact, the condition that a contact~$j$ must be detected by all processes
whose subdomains intersect with the hull intersection
$\mathcal{H}_{j_1} \cap \mathcal{H}_{j_2}$ is sufficient if the intersection of
the hull intersection and the domain is non-empty. This condition can be
fulfilled by the following requirement:
\begin{requirement}
\label{req:1}
A particle~$i$ must be instantiated not only
on the parent process but also on all processes whose subdomains intersect with
the particle's hull.
\end{requirement}
These additional instantiations shall be termed
\emph{shadow copies} in the following.
They must be kept in synchronization
with the original instantiation on the parent process. In order to agree upon
the detecting process responsible for treating the contact without
communication a rule is needed. Here, the statement that a process is
responsible for \emph{treating} a contact refers to the responsibility of the
process for executing the relaxation of the respective contact in
Alg.~\ref{alg:sdnbgs}. The typical choice for this rule 
requires that
the process whose subdomain contains the point of contact 
is put in charge 
to treat the
contact~\cite{shojaaee12}.

However, this rule only works if the process whose
subdomain contains the point is able to detect the contact. 
This is only
guaranteed if the point of contact is
located within the hull intersection. Also, if the point of contact is located
outside of the domain $\Omega$,
then no process will treat it.

A more intricate drawback
of this approach is that it can fail in case of 
periodic boundary conditions: If the
contact point is located near the periodic boundary, the periodic image of the
contact point will be detected at the other end of the simulation box. Due to
the shifted position of the contact point image and the limited numerical
precision, the subdomains can no longer consistently decide the subdomain
affinity.

A more robust rule to determine the subdomain affinity can be established by fulfilling the
following requirement:
\begin{requirement}
\label{req:2}
All shadow copy holders of a particle maintain a complete
list of all other shadow copy holders and the parent process of that particle.
\end{requirement}
Then each process
detecting a contact can determine the list of all processes detecting
that very same contact, which is the list of all processes with an instantiation
of both particles involved in the contact. This list is exactly the same
on all processes detecting the contact and is not prone to numerical errors.
The rule can then e.g.\ appoint the detecting process with smallest rank to
treat the contact. In order to enhance the locality of the contact treatment,
the rule should favor the particle parents if they are among the contact
witnesses. Any such rule defines a partitioning of the contact set $\contactset$.
Let~$\contactset_p$ be the set of all contacts treated by
process~$p \in \procset$. Then process~$p$ instantiates
all contacts~$j \in \contactset_p$.

\subsection{Accumulator and Correction Variables}
\label{sec:accumulator_and_correction_variables}

The contact relaxations in Alg.~\ref{alg:sdnbgs} exhibit sums with non-local data
dependencies. In the following, the redundant evaluation of these sums is
prevented by introducing accumulator variables and the non-local data
dependencies are reduced by introducing correction variables.

The relaxation of a contact~$j$ depends on the data of the state variables of
both particles $(j_1, j_2)$ involved in the contact, their constants and shape
parameters, as can be seen by inspecting \eqref{eq:nonpenetration}, \eqref{eq:coulomb} and the definitions
of the terms appearing therein. All of these quantities are instantiated on the detecting
process, either as a shadow copy or as an original instance. The
contact variables of contact~$j$ (location, signed distance and the contact frame)
are also required. They are available on the detecting process since they
result from the positions, orientations, and the shape parameters of the
particles $(j_1, j_2)$ in the contact detection. Furthermore,
the force and torque terms from \eqref{eq:wrench} acting on these particles
additionally depend on the locations $\contactpos{l}$ and reaction approximations $\contactforceapprox{l}^{(k,j)}$ of all other
contacts~$l$ involving one of the particles $(j_1, j_2)$. Neither the locations
nor the reaction approximations of these contacts are necessarily available on
the process treating contact~$j$. To rectify this deficiency, one can introduce
contact shadow copies so that location and reaction approximation can be
mirrored at every instantiation of both particles involved in the contact.
However, the organisational overhead of contact shadow copies can be circumvented.
It is not necessary that the process treating the contact evaluates all
the wrench contributions to the particles involved in the contact. Instead,
parts of the wrench contribution sum can be evaluated on the processes actually
treating the remote contacts and can subsequently be communicated:
\begin{equation*}
	\shrinkeqnnew{1}{
		\begin{split}
			\begin{pmatrix} \force{i}(\contactforce{}) \\ \torque{i}(\contactforce{}) \end{pmatrix}
			& = \begin{pmatrix} \forceext{i} \\ \torqueext{i} \end{pmatrix} + \sum_{\substack{l \in \contactset\\l_1 = i}} \begin{bmatrix} \mat{1} \\ (\contactpos{l} - \pos{i})^\cross \end{bmatrix} \contactforce{l} - \sum_{\substack{l \in \contactset\\l_2 = i}} \begin{bmatrix} \mat{1} \\ (\contactpos{l} - \pos{i})^\cross \end{bmatrix} \contactforce{l} \\
			& = \begin{pmatrix} \forceext{i} \\ \torqueext{i} \end{pmatrix} + \sum_{p \in \procset} \underbrace{\left( \sum_{\substack{l \in \contactset_p\\l_1 = i}} \begin{bmatrix} \mat{1} \\ (\contactpos{l} - \pos{i})^\cross \end{bmatrix} \contactforce{l} - \sum_{\substack{l \in \contactset_p\\l_2 = i}} \begin{bmatrix} \mat{1} \\ (\contactpos{l} - \pos{i})^\cross \end{bmatrix} \contactforce{l} \right)}_{\text{wrench contribution $(\force{i,p}\ \torque{i,p})^\transp$ to particle~$i$ from process~$p$}} \\
		\end{split}
	}
\end{equation*}

The total wrench on particle~$i$ can also be expressed in terms of the total
wrench on particle~$i$ at the beginning of iteration~$k$:
\begin{equation*}
	\shrinkeqnnew{1}{
		\begin{split}
			\begin{pmatrix} \force{i}(\contactforce{}) \\ \torque{i}(\contactforce{}) \end{pmatrix}
			= \begin{pmatrix} \force{i}(\contactforce{}^{(k)}) \\ \torque{i}(\contactforce{}^{(k)}) \end{pmatrix}
			+ \sum_{p \in \procset} & \left( \sum_{\substack{l \in \contactset_p\\l_1 = i}} \begin{bmatrix} \mat{1} \\ (\contactpos{l} - \pos{i})^\cross \end{bmatrix} (\contactforce{l} - \contactforce{l}^{(k)}) \right. \\
			& - \left. \sum_{\substack{l \in \contactset_p\\l_2 = i}} \begin{bmatrix} \mat{1} \\ (\contactpos{l} - \pos{i})^\cross \end{bmatrix} (\contactforce{l} - \contactforce{l}^{(k)}) \right)
		\end{split}
	}
\end{equation*}

When relaxing the contact~$j$ in iteration~$k$ of the subdomain NBGS, the wrench on particle~$i \in \set{j_1, j_2}$ is
evaluated with the reaction approximation $\contactforceapprox{}^{(k,j)}$ as parameter.
Since the subdomain NBGS respects the subdomain affinity of the
contacts, the remote wrench contributions to particle~$i$ cancel out, and just
the total wrench on particle~$i$ from the last iteration is needed in addition
to corrections stemming from contacts that were already relaxed by the same
process.
\begin{equation*}
	\shrinkeqnnew{1}{
		\begin{split}
			\begin{pmatrix} \force{i}(\contactforceapprox{}^{(k,j)}) \\ \torque{i}(\contactforceapprox{}^{(k,j)}) \end{pmatrix}
			= \begin{pmatrix} \force{i}(\contactforce{}^{(k)}) \\ \torque{i}(\contactforce{}^{(k)}) \end{pmatrix}
			& + \sum_{\substack{l \in \contactset_{s_c(j)} \\ l < j \\ l_1 = i}} \begin{bmatrix} \mat{1} \\ (\contactpos{l} - \pos{i})^\cross \end{bmatrix} (\contactforce{l}^{(k+1)} - \contactforce{l}^{(k)}) \\
			& - \sum_{\substack{l \in \contactset_{s_c(j)} \\ l < j \\ l_2 = i}} \begin{bmatrix} \mat{1} \\ (\contactpos{l} - \pos{i})^\cross \end{bmatrix} (\contactforce{l}^{(k+1)} - \contactforce{l}^{(k)})
		\end{split}
	}
\end{equation*}

Our implementation instantiates variables on process~$p$ for the reaction
approximations $\contactforce{}^{[p]} \in \R^{3\card{\contactset_p}}$ of all
contacts treated by process~$p$. Any updates to the reaction approximations
occur in place. Furthermore, an implementation can instantiate accumulator
variables~$\force{}^{[p]}, \torque{}^{[p]} \in \R^{3 \card{\bodyset_p}}$
on process~$p$ for the wrenches from the last iteration of all instantiated
particles (shadow copies and original instances), where $\bodyset_p$ contains
the indices of all shadow copies and original instances instantiated on process~$p$.
This set is partitioned into $\bodyset_{p,local}$ and $\bodyset_{p,shadow}$,
containing the indices of the original instances and the shadow copies respectively.

Instead of evaluating the wrench contribution sums each time when calculating
the total wrench on particle~$i$ anew, the contributions can be accumulated as the
contacts are relaxed. For that purpose, implementations can instantiate
corrections variables $\delta \force{}^{[p]} \in \R^{3 \card{\bodyset_p}}$ and
$\delta \torque{}^{[p]} \in \R^{3 \card{\bodyset_p}}$. Then, after line~\ref{line:update} of
Alg.~\ref{alg:sdnbgs}, these wrench corrections can be updated by assigning
\begin{equation*}
	\shrinkeqnnew{1}{
		\begin{split}
			& \dvect{\delta \force{j_1}^{[s_c(j)]}}{\delta \torque{j_1}^{[s_c(j)]}} \gets \dvect{\delta \force{j_1}^{[s_c(j)]}}{\delta \torque{j_1}^{[s_c(j)]}} + \begin{bmatrix} \mat{1} \\ (\contactpos{j} - \pos{j_1})^\cross \end{bmatrix} (\contactforce{j}^{(k+1)} - \contactforce{j}^{(k)}), \\
			& \dvect{\delta \force{j_2}^{[s_c(j)]}}{\delta \torque{j_2}^{[s_c(j)]}} \gets \dvect{\delta \force{j_2}^{[s_c(j)]}}{\delta \torque{j_2}^{[s_c(j)]}} - \begin{bmatrix} \mat{1} \\ (\contactpos{j} - \pos{j_2})^\cross \end{bmatrix} (\contactforce{j}^{(k+1)} - \contactforce{j}^{(k)}).
		\end{split}
	}
\end{equation*}

The evaluation of the total wrench on particle~$i$ in line~\ref{line:solve} of
Alg.~\ref{alg:sdnbgs} when relaxing contact~$j$ in iteration~$k$ becomes
\begin{equation*}
	\begin{pmatrix} \force{i}(\contactforceapprox{}^{(k,j)}) \\ \torque{i}(\contactforceapprox{}^{(k,j)}) \end{pmatrix} = \begin{pmatrix} \force{i}^{[s_c(j)]} \\ \torque{i}^{[s_c(j)]} \end{pmatrix} + \dvect{\delta \force{i}^{[s_c(j)]}}{\delta \torque{i}^{[s_c(j)]}},
\end{equation*}
that is the sum of the accumulator and the correction variables.

At the end of each iteration the wrench corrections for each body have to be
reduced and added to the accumulated wrench from the last iteration. This can be
performed in two message exchanges. In the first message exchange each process
sends the wrench correction of each shadow copy to its parent process. Then
each process sums up for each original instance all wrench corrections obtained
from the shadow copy holders, its own wrench correction, and the original instance's
accumulated wrench. Subsequently, the updated accumulated wrench of each original
instance is sent to the shadow copy holders in a second message-exchange
communication step. The wrench corrections are then reset everywhere.

The accumulated wrenches $\force{}^{[p]}$, $\torque{}^{[p]}$ are initialized on each process~$p$ before line~\ref{line:loopheader} in Alg.~\ref{alg:sdnbgs}
to
\begin{equation*}
	\dvect{\force{i}^{[p]}}{\torque{i}^{[p]}} \gets \begin{pmatrix} \forceext{i} \\ \torqueext{i} \end{pmatrix} \quad \forall i \in \bodyset_p
\end{equation*}
unless the initial solution is chosen to be non-zero. The wrench corrections are
initially set to $\vec 0$. If the external forces and torques are not known on
each process or are scattered among the processes having instantiated the particles,
the initialization requires another two message exchanges, as they are necessary at
the end of each iteration.

An alternative to storing accumulated wrenches and wrench corrections is to store accumulated velocities
and velocity corrections.
In that case, a process~$p$ instantiates variables
$\linvel{}^{[p]}$, $\angvel{}^{[p]}$, $\delta \linvel{}^{[p]}$, $\delta \angvel{}^{[p]} \in \R^{3 \card{\bodyset_p}}$.
The accumulated velocities are set to $\linvel{i}'(\contactforce{}^{(k)})$ and
$\angvel{i}'(\contactforce{}^{(k)})$ for all $i \in \bodyset_p$ in each iteration.
They are initialized and updated accordingly. The velocity
corrections are initialized and updated analogously to the
wrench corrections. Hereby, the velocity variables can be updated in place.
In the classic NBGS no wrench or velocity correction variables would
be necessary, but the corrections could be added to the velocity variables
right away which is similar to the approach suggested by Tasora et al.\ in \cite{tasora11}.

\subsection{Nearest-Neighbor Communication}
\label{sec:nearest_neighbor_communication}

In the following we describe how the strict locality of particle interactions can
be used to optimize the parallel communication and synchronization by
exchanging messages only between nearest neighbors. So far the shadow copies
can be present on any process, and the corrections in
the summation over wrench or velocity corrections can originate from a long list
of processes. However, by requiring that the particle hulls do not extend past
any neighboring subdomains, all message exchanges can be reduced to
nearest-neighbor communications. Let
\begin{equation*}
	\shrinkeqnnew{1}{
		\neighborset_p = \setprop{q \in \procset \setminus \set{p}}{\inf\setprop{\norm{\vec y_p - \vec y_q}_2}{\vec y_p \in \Omega_p, \vec y_q \in \Omega_q} = 0}
	}
\end{equation*}
be the set of process indices in direct neighborhood of process~$p$'s subdomain,
and let
\begin{equation*}
	\shrinkeqnnew{1}{
		l_{dd} = \min_{p \in \procset} \operatorname{inf} \setprop{\norm{\vec y_p - \vec y_q}_2}{\vec y_p \in \Omega_p, \vec y_q \in \bigcup_{q \in \procset \setminus (\neighborset_p \cup \set{p})} \Omega_q},
	}
\end{equation*}
be the shortest distance from a point inside a subdomain to a non-nearest neighbor.
Then the condition
\begin{equation}
	\label{eq:ldd}
	\overline{r}_i + \norm{\linvel{i}(t)}_2 \dt + \tau < l_{dd} \quad \forall i \in \bodyset
\end{equation}
ensures in the first approximation that no hull extends past neighboring
subdomains. This immediately defines a hard upper limit of
$\overline{r}_i < l_{dd} - \tau$ for the bounding radius and thus for the size of all
objects. Furthermore, given the particle shapes, velocities, and safety margins, the condition
defines an upper limit for the time-step length. The introduction of condition~\eqref{eq:ldd}
entails that on a process~$p$ only the description of the subdomains within
$\Omega_p + \setprop{\vec y \in \R^3}{\norm{\vec y}_2 \leq l_{dd}}$ needs to be
available, meaning that the description of non-nearest-neighbor subdomains can
be dispensed with, and that the description of nearest-neighbor subdomains do not
have to be correct outside of the $l_{dd}$-surrounding of $\Omega_p$. This leads
to a \emph{localized} description of the domain partitioning on each process,
describing the surrounding subdomains only.

Typically, the size limit stemming from \eqref{eq:ldd} is not a problem
for the particles of the granular matter themselves, but very well for boundaries
or mechanical parts the granular matter interacts with. However, the
number of such enlarged bodies is typically significantly smaller than and independent of
the number of small-sized particles, suggesting that they can be treated
globally. Let $\bodyset_{global}$ be the set of all body indices exceeding
the size limit. These bodies will be referred to as being global in the
following. All associated state variables and constants shall be instantiated
on all processes and initialized equally. The time-integration of these global
bodies then can be performed by all processes equally. If a global body~$i$
has infinite inertia ($\mass{i} = \infty$ and $\inertia{ii}^0 = \infty \mat{1}$),
such as a stationary wall or a non-stationary vibrating plate, the body
velocities are constant, and no wrenches need to be communicated. Global
bodies having a finite inertia can be treated by executing an all-reduce
communication primitive whenever reducing the wrench or velocity corrections of
the small-sized bodies. Instead of only involving neighboring processes, the all-reduce operation
sums up the corrections for each global body with finite inertia from all
processes and broadcasts the result, not requiring any domain partitioning
information.

\subsection{Time-Integration and Synchronization Protocol}
\label{sec:time-integration_and_synchronization}

Having solved the contact problem $\vec{F}(\contactforce{}) = \vec 0$ by
Alg.~\ref{alg:sdnbgs}, the time-integration defined in \eqref{eq:integrationscheme} needs to be
performed. If the NBGS implementation uses velocity accumulators, the
integrated velocities are at hand after the final communication of
the velocity corrections. If instead the NBGS implementation uses wrench
accumulators, the wrenches are at hand, and the velocities of all local bodies
can be updated immediately.

Subsequently, the time-integration of the
positions can take place. Updating a body's position or orientation effects
that the list of shadow copy holders changes since the intersection hull
possibly intersects with different subdomains. Also, the body's center of mass
can move out of the parent's subdomain. In order to restore
the fulfillment of the requirements \ref{req:1} and \ref{req:2}, a process must
determine the new list of shadow copy holders and the new parent process for
each local body after the position update. 
Shadow copy holders must be informed when such shadow copies become obsolete and must be removed.
Analogously, processes must be notified when
new shadow copies must be added to their state.
In this case
copies
of the corresponding state variables, constants, list of shadow copy holder indices,
and index of the parent process must be transmitted.

All other shadow copy holders must obtain the new state variables,
list of shadow copy holder indices, and index of the parent process.
Hereby, the condition from \eqref{eq:ldd} guarantees that all communication
partners are neighbors. All information can be propagated in a single
aggregated nearest-neighbor message-exchange. The information
should be communicated explicitly and should not be derived implicitly, in order
to avoid inconsistencies. 
This is essential to guarantee a safe determination of the
contact treatment responsibilities as well as time-integration responsibilities.

Our implementation of the synchronization protocol makes use of separate
containers for storing shadow copies and original instances in order to be able
to enumerate these different types of bodies with good performance.
Both containers support efficient insertion, deletion and lookup
operations for handling the fluctuations and updates of the particles efficiently.
Furthermore, the determination of the new list of shadow copy holders involves intersection
tests between intersection hulls of local bodies and neighboring subdomains
as requirement~\ref{req:1} explains in Sect.~\ref{sec:shadow_copies}.
However, determining the minimal set of shadow copy holders is not necessary.
Any type of bounding volumes can be used to ease intersection testing.
In particular bounding spheres either with tightly fitting bounding radii
$\overline{r}_i + h_i(t)$ or even with an overall bounding radius
$\max_{i \in \bodyset} \overline{r}_i + h_i(t)$ as proposed by Shojaaee
et al.\ in \cite{shojaaee12} are canonical. Concerning the geometry of the
subdomains at least the subdomain closures can be used for intersection testing.
In our implementation we chose to determine almost minimal sets of shadow copy
holders by testing the intersections of the actual hull geometries of the particles
with the closures of the subdomains. This reduces the number of shadow copies and
thus the overall communication volume in exchange for more expensive intersection tests.

\subsection{Summary}
\label{sec:parallelization_summary}

Alg.~\ref{alg:simulate-time-step} summarizes the steps that need to be
executed on a process~$p$ when time-integrating the system for a single time
step~$\dt$ in parallel.
\begin{algorithm}
	\begin{algorithmic}[1]
		\Procedure{simulateTimeStep}{}
			\State $\contactset_{\mathit{p,bp}} = \Call{broadPhaseCollisionDetection}{}$
			\State $\contactset_{\mathit{p,np}} = \Call{narrowPhaseCollisionDetection}{\contactset_{\mathit{p,bp}}}$
			\State $\contactset_p = \Call{filterContacts}{\contactset_{\mathit{p,np}}}$
			\State \Call{initializeAccumulatorAndCorrectionVariables}{}
			\State $k \gets 0$
			\State $\contactforce{}^{[p]} \gets \vec 0$
			\While{convergence criterion not met}
				\For{$j \gets 1\ \algorithmicto\ \numcontacts \land j \in \contactset_p$}
					\State $\contactforce{j}^{[p]} \gets \omega \vec F_j^{-1}(\vec 0, \contactforce{\overbar{j}}^{[p]}) + (1 - \omega) \contactforce{j}^{[p]}$
				\EndFor
				\State \Call{reduceCorrections}{}
				\State $k \gets k + 1$
			\EndWhile
			\State \Call{integrateStateVariables}{}
			\State \Call{synchronize}{}
		\EndProcedure
	\end{algorithmic}
	\caption{A single time step of the simulation on process~$p$.\label{alg:simulate-time-step}}
\end{algorithm}
The algorithm requires that all shadow copies are instantiated
on all subdomains their hull intersects with. Furthermore, the shadow copies
must be in sync with the original instance, and the global bodies must also
be in sync to each other. The positions of all local bodies must be located
within the local subdomain. The time step proceeds by executing the
broad-phase contact detection which uses the positions, orientations, shapes,
hull expansion radii, and possibly 
information from previous time steps, in
order to determine a set of contact candidates (body pairs) $\contactset_{p,bp}$ on
process~$p$ in near-linear time.

Then, in the narrow-phase contact detection,
for all candidates the contact location, associated contact frame, and signed
contact distance is determined if the hulls actually intersect. Finally, this set
of detected contacts $\contactset_{p,np}$ needs to be filtered according
to one of the rules presented above, resulting in $\contactset_p$, the set of
contacts to be treated by process~$p$. Before entering the iteration of the
subdomain NBGS, the accumulator, correction, and contact reaction variables must
be initialized. The initialization of the accumulator variables requires
an additional reduction step if the external forces or torques cannot be
readily evaluated on all processes.

Each iteration of the subdomain NBGS on process~$p$ involves a sweep over all contacts
to be treated by the process. The contacts are relaxed by a suitable one-contact
solver. The $\overbar{j}$ indexing indicates that such a solver typically
needs to evaluate the relative contact velocity under the assumption that
no reaction acts at the contact~$j$. This can be achieved by
subtracting out the corresponding part from the accumulator variables. The
weighted relaxation result is then stored in place. The update of the wrench
or velocity correction variables
is not explicitly listed. After the sweep the wrench or velocity corrections
are sent to the respective parent process and summed up per body including the
accumulator variables. Then the accumulator variables are redistributed to the
respective shadow copy holders in a second message-exchange step.

After
a fixed number of iterations or some prescribed convergence criterion is met,
the time step proceeds by executing the time-integration for each local body.
The 
changes of the state variables must
then be synchronized in a
final message-exchange step, after which the preconditions of the next time step
are met. Any user intervention taking place between two time steps needs to
adhere to these requirements.

\section{Experimental Validation of Scalability}
\label{sec:scaling_experiments}

This section aims to assess the scalability of the parallelization design
presented in Sect.~\ref{sec:parallelization_design} as we implemented it
in the \pe{} which is an open-source software framework for massively parallel
simulations of rigid bodies~\cite{iglberger2010massively,iglberger11}.
The implementation is based on
velocity accumulators and corrections, as introduced in
Sect.~\ref{sec:accumulator_and_correction_variables}.
The accumulator initialization performs an additional initial correction
reduction step in all experiments.

In Sect.~\ref{sec:weak_and_strong_scalability} the idea behind
weak- and strong-scaling experiments is explained before presenting
the test problems for which those experiments are executed in
Sect.~\ref{sec:test_problems}. The scaling experiments are performed
on three clusters whose properties are summarized and compared in
Sect.~\ref{sec:test_machines}. Sect.~\ref{sec:time-step_profiles}
points out the fundamental differences in the scalability
requirements of the two test problems. Finally, in Sect.~\ref{sec:weak-scaling_results}
the weak-scaling and in Sect.~\ref{sec:strong-scaling_results} the strong-scaling
results are presented for each test problem and cluster.

\subsection{Weak and Strong Scalability}
\label{sec:weak_and_strong_scalability}

To demonstrate the scalability of the algorithms and their implementation,
we perform 
weak-scaling experiments, where the problem size is chosen directly proportional to
the number of processes 
and such that the load per
process stays constant. 
Thus, if ideal scaling were achieved, 
the time to solution 
would stay constant.
Let~$t_p$ be the time to solution on $p$~processes, then the parallel
efficiency~$e_{p,ws}$ in a weak-scaling experiment is defined to be
\begin{equation*}
	e_{p,ws} = \frac{t_1}{t_p}.
\end{equation*}


In strong-scaling experiments, in contrast, the problem size
is kept constant, effecting a decreasing work load per process when increasing
the number of processes. 
Thus, ideally the time
to solution on $p$~processes should be 
reduced by $p$ in comparison
to the time to solution on a single process. The 
speedup~$s_p$ on
$p$~processes is defined to be
\begin{equation*}
	s_p = \frac{t_1}{t_p}.
\end{equation*}
The parallel efficiency~$e_{p,ss}$ in a strong-scaling experiment is then the
fraction of the ideal speedup actually achieved
\begin{equation*}
	e_{p,ss} = \frac{s_p}{p} = \frac{t_1}{p t_p}.
\end{equation*}

Sometimes speedup and parallel efficiency are also stated with respect to a
different baseline, that is, a single central processing unit~(CPU) or a single node rather than a
single hardware thread or core -- the principle remains the same.
The parallel efficiency
in a weak- and strong-scaling context
is a simple performance metric
that will serve in the
following to assess the quality of the parallelization.

\subsection{Test Problems}
\label{sec:test_problems}

The scalability of the parallelization algorithm as it is implemented in the
\pe{} framework is validated based on two fundamentally different families of
test problems. Sect.~\ref{sec:granular_gas} describes a family of dilute granular gas
setups whereas Sect.~\ref{sec:hcps} describes a family of hexagonal close
packings of spheres corresponding to structured and dense setups. We chose
these setups because their demands towards the implementation vary considerably which will
be analyzed in detail in Sect.~\ref{sec:time-step_profiles}.

\subsubsection{Granular Gas}
\label{sec:granular_gas}

Granular
material attains a gaseous state when sufficient energy is brought into the
system, for example by vibration. Consequently, granular gases feature a
low solid volume fraction and are dominated by binary collisions. When
the energy supply 
ceases, the system cools down due 
to dissipation in the 
collisions. 
Granular gases are not only observed in laboratory
experiments, but appear naturally for example in planetary rings~\cite{spahn01} and in technical applications such as granular dampers~\cite{kollmer13}. These
systems in general exhibit interesting effects like the inelastic collapse~\cite{mcnamara94} or other
clustering effects as they e.g.\ can be observed in the Maxwell-demon experiment~\cite{weele01}.

As initial
conditions, a rectangular domain with confining walls is chosen. The
domain contains a prescribed number of 
non-spherical particles arranged in a Cartesian
grid. Random initial velocities are assigned to the particles with up
to $\unitfrac[\frac{2\sqrt{3}}{10}]{\meter}{\second} \approx \unitfrac[0.35]{\meter}{\second}$.
The 
particles are composed of two to four spheres of varying
radius in the range $\intervalCC{\unit[0.6]{\centi\meter}}{\unit[0.8]{\centi\meter}}$, arranged at the boundary of a bounding sphere with a diameter
of~$\unit[1]{\centi\meter}$. The distance between the centers of two granular
particles along each spatial dimension is~$\unit[1.1]{\centi\meter}$, amounting
to a solid volume fraction of 23\% on average.
In \cite{iglberger2010massively} almost the same family of setups served as a scalability
test problem. However, there the granular gases had a solid volume fraction of
3.8\% on average.
In order
to 
test a higher collision frequency, a more dense granular gas was chosen here.
The system is simulated for $\unit[\frac{1}{10}]{\second}$, and the time step is
kept constant at $\unit[100]{\micro\second}$, resulting in 1\,000~time steps in
total. Since the contacts are dissipative and no energy is added, the system is
quickly cooling down. The coefficient of friction is~$0.1$ for any contact.

For this test problem, the
subdomain NBGS solver requires a slight underrelaxation in order to prevent
divergence. Using an underrelaxation parameter
of~0.75 
produces good results.
For binary collisions, a single iteration of the solver would suffice,
but because particles cluster due to the inelastic contacts, more iterations are
required. This could be determined by a dynamic stopping criterion, but in the
scenario presented here it was found to be more efficient to perform a
fixed number of 10~iterations.

For particle simulations, the work load strongly depends on the number of
particles and contacts. 
For the weak-scaling experiments, each
process is responsible for a rectangular subdomain, 
initially %
containing a
fixed number of particles arranged in a Cartesian grid. For the strong-scaling
experiments, the total number of particles in x-, y-, and z-dimension
should be divisible by the number of processes in x-, y-, and z-dimension
that is %
used in the experiment. 
With this arrangement the initial load is perfectly balanced.
Statistically, the load, that is the number of 
particles and contacts per subdomain,
remains balanced if the subdomains are large enough, and clustering %
effects have not yet 
progressed too far.
The 
duration of the simulation was chosen such that
the load 
remains well balanced 
throughout.

\subsubsection{Hexagonal Close Packing of Spheres}
\label{sec:hcps}

This setup aims to assess the scalability of the parallelization
for dense granular setups. To demonstrate the scalability, the initial setup should be
easily and efficiently generateable for arbitrary problem sizes
and should feature a good load balance over a longer period of time.
Hence, a hexagonal close packing of equal spheres was chosen, for
which simple formulas for the position of the spheres are available.
The packing density is known to be $\frac{\pi}{3\sqrt{2}} \approx 74.0\%$.
According to the Kepler conjecture, a hexagonal close packing is
the densest possible packing of spheres. To avoid load imbalances
at the boundaries, a domain is chosen that is periodic in the x- and y-dimension.
In z-dimension the packing is confined by walls that are
in direct contact with the spheres on both sides. Assuming an even
number of particles in y-direction, the number of contacts is permanently
$n_x n_y (6n_z - 1)$ for $n_x \times n_y \times n_z$ particles.
The domain is decomposed
in x- and y-dimensions only. The objects are subject to gravity.
However, the gravity is tilted in the x-z-plane such that the setup corresponds to
a ramp inclined by~30° including a lid. The magnitude of gravity is
$\unitfrac[9.81]{\meter}{\square\second}$. The time step is $\unit[10]{\micro\second}$ constantly.
The radii of the particles are $\unit[1]{\milli\meter}$, and their density is $\unitfrac[2.65]{\gram}{\centi\meter\cubed}$.
All particles get an initial downhill velocity of $\unitfrac[10]{\centi\meter}{\second}$.
The coefficient of friction is~$0.85$ constantly for any contact. The high coefficient of friction
causes a slip-stick transition shortly after the simulation begins.
As in the granular gas setups the subdomain NBGS uses an underrelaxation of 0.75.
The solver unconditionally performs~100 iterations in each time step. This
intentionally disregards that the iterative solver converges faster for smaller problems.
A multigrid solver could possibly remedy the dependence on the problem size, but
the successful construction of such a solver needs substantial further research.

\subsection{Test Machines}
\label{sec:test_machines}

In the following all test machines are presented. Tab.~\ref{tab:machines} summarizes
the basic information.
\begin{table*}
\centering
\begin{tabular}{llll}
\toprule
cluster name                                            & Emmy                                                 & SuperMUC                             & Juqueen \\
\midrule
computing centre                                        & \mrcell{Regional Computing Centre in\\ Erlangen (RRZE), Germany}  & \mrcell{Leibniz Supercomputing\\ Centre (LRZ), Germany}  & \mrcell{Jülich Supercomputing\\ Centre (JSC), Germany} \\
best TOP 500 ranking                                    & -                                                    & 4th (June 2012)                                 & 5th (November 2012) \\
peak performance in $\unitfrac[]{\peta\flop}{\second}$  & 0.23                                                 & 3.2                                             & 5.9 \\
number of nodes                                         & 560                                                  & 9\,216                                          & 28\,672 \\
number of sockets                                       & 2                                                    & 2                                               & 1 \\
name of CPU                                             & Intel Xeon~E5-2660~v2                                & Intel Xeon~E5-2680                              & IBM PowerPC~A2 \\
clock rate in $\unit[]{\giga\hertz}$                    & 2.2                                                  & 2.7                                             & 1.6 \\
number of cores per CPU                                 & 10                                                   & 8                                               & 16 \\
number of threads per core                              & 2                                                    & 2                                               & 4 \\
total RAM in $\unit[]{\tebi\byte}$                      & 35                                                   & 288                                             & 448 \\
interconnection fabric                                  & Infiniband QDR                                       & \mrcell{Infiniband QDR/\\Infiniband FDR 10}     & BlueGene/Q \\
network topology                                        & non-blocking tree                                    & \mrcell{non-blocking tree/\\4:1 pruned tree}    & 5D~torus \\
\bottomrule
\end{tabular}
\caption{The test machines used for performing the weak- and strong-scaling experiments.\label{tab:machines}}
\end{table*}
The Emmy cluster is located at the Regional Computing Centre in Erlangen~(RRZE) in Germany which
is associated to the Friedrich-Alexander-Universität Erlangen-Nürnberg. The
cluster comprises 560~compute nodes. Each node has a dual-socket board equipped
with two Xeon~E5-2660~v2 processors. Each processor has 10~cores clocked at
$\unit[2.2]{\giga\hertz}$. The processors offer 2-way~simultaneous multithreading~(SMT).
The peak performance of the cluster is\linebreak $\unitfrac[0.23]{\peta\flop}{\second}$.
Each node 
is equipped with $\unit[64]{\gibi\byte}$ of random access memory~(RAM). The
cluster features a fully non-blocking Infiniband interconnect with quad data rate (QDR)
and $4 \times$ link aggregation, resulting in a bandwidth of $\unitfrac[40]{\giga\bit}{\second}$ per
link and direction.
In all experiments on the Emmy cluster, each core is associated with a subdomain
since preliminary tests showed that we could not take advantage of the SMT
features by associating each hardware thread with a subdomain.
The Emmy cluster has the smallest peak performance among the
test machines and was never among the 500 world's largest commercially available
supercomputers. However, it is the only machine with the largest non-blocking tree
network topology and the largest amount of RAM per core.

The second test machine is the SuperMUC supercomputer which is located at the
Leibniz Supercomputing Centre~(LRZ) in Germany and was best 
ranked on the 4th place of the TOP~500 list in June~2012.
The cluster is subdivided into multiple islands. The majority of the compute power
is contributed by the 18~thin-node islands. Each thin-node island consists of
512~compute nodes (excluding four additional spare nodes) connected to a fully
non-blocking 648 port FDR10 Infiniband switch with $4\times$ link aggregation,
resulting in a bandwidth of $\unitfrac[40]{\giga\bit}{\second}$ per
link and direction. Though QDR and FDR10 use the same signaling rate, the effective
data rate of FDR10 is more than 20\% higher since it uses a more efficient
encoding of the transmitted data. The islands' switches are each connected via
126~links to 126~spine switches. This results in a blocking switch-topology.
Thus, if e.g.\ all nodes within an island send to nodes located in another
island, then the 512~nodes have to share 126 links to the spine switches,
effecting that the bandwidth is roughly one quarter of the bandwidth that
would be available in an overall non-blocking switch-topology.
Each (thin) compute node has two sockets, each equipped
with an Intel Xeon E5-2680 processor having 8~cores clocked at $\unit[2.7]{\giga\hertz}$.
The processors support 2-way~SMT. In the following, as in the case of the Emmy cluster, each core
is associated with a single subdomain. The peak performance of the cluster is stated to be
$\unitfrac[3.2]{\peta\flop}{\second}$. Each node offers $\unit[32]{\gibi\byte}$
of RAM, summing up to~$\unit[288]{\tebi\byte}$ in total. The SuperMUC supercomputer
has an interesting blocking tree network-topology and the processors with the
highest clock rate among the processors in the test machines.

The third test machine is the Juqueen supercomputer which is located at the
Jülich Supercomputing Centre~(JSC) in Germany
and was best ranked on the 5th place of the TOP~500 list in November~2012.
The cluster is a BlueGene/Q system with 28\,672~compute nodes since~2013~\cite{gilge13,wautelet14}. Each node
features a single IBM PowerPC~A2 processor having 18~cores clocked at
$\unit[1.6]{\giga\hertz}$, where only 16~cores are available for computing.
The processors support 4-way~SMT. The Juqueen supercomputer is the only machine,
where we decided to associate each hardware thread with a subdomain in the
scaling experiments. The machine's peak performance is $\unitfrac[5.9]{\peta\flop}{\second}$.
Each node offers $\unit[16]{\gibi\byte}$
of RAM, summing up to~$\unit[448]{\tebi\byte}$ in total. The interconnect fabric
is a 5D~torus network featuring a bandwidth of $\unitfrac[16]{\giga\bit}{\second}$
per link and direction~\cite{chen12}. The Juqueen supercomputer is the machine
with the highest peak performance, the largest number of cores and threads and
the only machine among our test machines with a torus interconnect.

Tab.~\ref{tab:partitions} presents a summary of the domain partitionings used for
the scaling experiments on the various clusters.
\begin{table*}
	\centering
	\subfloat[Domain partitionings used on the Emmy cluster.\label{tab:partitions_emmy}]{%
		\begin{tabular}{llllllllllllllllll}
			\toprule
			\multirow{3}{*}{1D} & nodes   & $\frac{1}{20}$ & $\frac{2}{20}$ & $\frac{4}{20}$ & $\frac{8}{20}$ & $\frac{10}{20}$ & $\frac{16}{20}$ & 1  & 2  & 4  & 8   & 16  & 32  & 64     & 128    & 256    & 512     \\[3pt]
			                    & $p_x$   & 1              & 2              & 4              & 8              & 10              & 16              & 20 & 40 & 80 & 160 & 320 & 640 & 1\,280 & 2\,560 & 5\,120 & 10\,240 \\
			                    &         & \multicolumn{16}{c}{\tikzmark{e1}\hfill\tikzmark{s1}\ weak-scaling granular gas\ \tikzmark{s2}\hfill\tikzmark{e2}}                                                          \\
			\midrule
			\multirow{6}{*}{2D} & nodes   & $\frac{4}{20}$ & $\frac{8}{20}$ & $\frac{10}{20}$ & $\frac{16}{20}$ & 1 & 2 & 4  & 8  & 16 & 32 & 64 & 128 & 256 & 512                                                                                    \\[3pt]
			                    & $p_x$   & 2              & 4              & 5               & 4               & 5 & 8 & 10 & 16 & 20 & 32 & 40 & 64  & 80  & 128                                                                                    \\
			                    & $p_y$   & 2              & 2              & 2               & 4               & 4 & 5 & 8  & 10 & 16 & 20 & 32 & 40  & 64  & 80                                                                                     \\
			                    &         & \multicolumn{14}{c}{\tikzmark{e3}\hfill\tikzmark{s3}\ weak-scaling granular gas\ \tikzmark{s4}\hfill\tikzmark{e4}}                                                                                        \\
			                    &         &                &                &                 &                 & \multicolumn{10}{c}{\tikzmark{e9}\hfill\tikzmark{s9}\ weak-scaling hexagonal close packing\ \tikzmark{s10}\hfill\tikzmark{e10}}     \\
			                    &         &                &                &                 &                 & \multicolumn{10}{c}{\tikzmark{e11}\hfill\tikzmark{s11}\ strong-scaling hexagonal close packing\ \tikzmark{s12}\hfill\tikzmark{e12}} \\
			\midrule
			\multirow{6}{*}{3D} & nodes   & $\frac{8}{20}$ & $\frac{16}{20}$ & 1 & 2 & 4 & 8 & 16 & 32 & 64 & 128 & 256 & 512                                                                       \\[3pt]
			                    & $p_x$   & 2              & 4               & 5 & 5 & 5 & 8 & 8  & 10 & 16 & 16  & 20  & 32                                                                        \\
			                    & $p_y$   & 2              & 2               & 2 & 4 & 4 & 5 & 8  & 8  & 10 & 16  & 16  & 20                                                                        \\
			                    & $p_z$   & 2              & 2               & 2 & 2 & 4 & 4 & 5  & 8  & 8  & 10  & 16  & 16                                                                        \\
			                    &         & \multicolumn{12}{c}{\tikzmark{e5}\hfill\tikzmark{s5}\ weak-scaling granular gas\ \tikzmark{s6}\hfill\tikzmark{e6}}                                      \\
			                    &         &                &                 & \multicolumn{10}{c}{\tikzmark{e7}\hfill\tikzmark{s7}\ strong-scaling granular gas\ \tikzmark{s8}\hfill\tikzmark{e8}} \\
			\bottomrule
		\end{tabular}
		\begin{tikzpicture}[overlay,remember picture]
			\foreach \i in {1,2,...,12}
				\draw[->|] ($(s\i.mid)$) -- ($(e\i.mid)$);
		\end{tikzpicture}
	}\hfill
	\subfloat[Domain partitionings used on the Juqueen supercomputer.\label{tab:partitions_juqueen}]{%
		\begin{tabular}{llllllllllllllllll}
			\toprule
			\multirow{5}{*}{2D} & nodes   & 1 & 2  & 4  & 8  & 16 & 32 & 64 & 128 & 256 & 512 & 1\,024 & 2\,048 & 4\,096 & 8\,192 & 16\,384 & 28\,672                                                    \\[3pt]
			                    & $p_x$   & 8 & 16 & 16 & 32 & 32 & 64 & 64 & 128 & 128 & 256 & 256    & 512    & 512    & 1\,024 & 1\,024  & 1\,024                                                     \\
			                    & $p_y$   & 8 & 8  & 16 & 16 & 32 & 32 & 64 & 64  & 128 & 128 & 256    & 256    & 512    & 512    & 1\,024  & 1\,792                                                     \\
			                    &         & \multicolumn{16}{c}{\tikzmark{e1}\hfill\tikzmark{s1}\ weak-scaling hexagonal close packing\ \tikzmark{s2}\hfill\tikzmark{e2}}                                \\
			                    &         &   &    &    &    &    & \multicolumn{8}{c}{\tikzmark{e3}\hfill\tikzmark{s3}\ strong-scaling hexagonal close packing\ \tikzmark{s4}\hfill\tikzmark{e4}} & & & \\
			\midrule
			\multirow{5}{*}{3D} & nodes   & 1 & 2 & 4 & 8 & 16 & 32 & 64 & 128 & 256 & 512 & 1\,024 & 2\,048 & 4\,096 & 8\,192 & 16\,384 & 28\,672                     \\[3pt]
			                    & $p_x$   & 4 & 8 & 8 & 8 & 16 & 16 & 16 & 32  & 32  & 32  & 64     & 64     & 64     & 128    & 128     & 128                         \\
			                    & $p_y$   & 4 & 4 & 8 & 8 & 8  & 16 & 16 & 16  & 32  & 32  & 32     & 64     & 64     & 64     & 128     & 128                         \\
			                    & $p_z$   & 4 & 4 & 4 & 8 & 8  & 8  & 16 & 16  & 16  & 32  & 32     & 32     & 64     & 64     & 64      & 112                         \\
			                    &         & \multicolumn{16}{c}{\tikzmark{e5}\hfill\tikzmark{s5}\ weak-scaling granular gas\ \tikzmark{s6}\hfill\tikzmark{e6}}         \\
			                    &         & \multicolumn{13}{c}{\tikzmark{e7}\hfill\tikzmark{s7}\ strong-scaling granular gas\ \tikzmark{s8}\hfill\tikzmark{e8}} & & & \\
			\bottomrule
		\end{tabular}
		\begin{tikzpicture}[overlay,remember picture]
			\foreach \i in {1,2,...,8}
				\draw[->|] ($(s\i.mid)$) -- ($(e\i.mid)$);
		\end{tikzpicture}
	}\hfill
	\subfloat[Domain partitionings used on the SuperMUC supercomputer.\label{tab:partitions_supermuc}]{%
		\begin{tabular}{llllllllllllllllll}
			\toprule
			\multirow{6}{*}{3D} & nodes   & 1 & 2 & 4 & 8 & 16 & 32 & 64 & 128 & 256 & 512 & 1\,024 & 2\,048 & 4\,096 & 8\,192                                         \\[3pt]
			                    & $p_x$   & 4 & 4 & 4 & 8 & 8  & 8  & 16 & 16  & 16  & 32  & 32     & 32     & 64     & 64                                             \\
			                    & $p_y$   & 2 & 4 & 4 & 4 & 8  & 8  & 8  & 16  & 16  & 16  & 32     & 32     & 32     & 64                                             \\
			                    & $p_z$   & 2 & 2 & 4 & 4 & 4  & 8  & 8  & 8   & 16  & 16  & 16     & 32     & 32     & 32                                             \\
			                    &         & \multicolumn{14}{c}{\tikzmark{e1}\hfill\tikzmark{s1}\ weak-scaling granular gas\ \tikzmark{s2}\hfill\tikzmark{e2}}         \\
			                    &         & \multicolumn{11}{c}{\tikzmark{e3}\hfill\tikzmark{s3}\ strong-scaling granular gas\ \tikzmark{s4}\hfill\tikzmark{e4}} & & & \\
			\bottomrule
		\end{tabular}
		\begin{tikzpicture}[overlay,remember picture]
			\foreach \i in {1,2,...,4}
				\draw[->|] ($(s\i.mid)$) -- ($(e\i.mid)$);
		\end{tikzpicture}
	}
	\caption{Summary of the domain partitionings used on all test clusters.\label{tab:partitions}}
\end{table*}
The number of nodes are
always a power of two except when using the whole machine or when performing
intra-node scalings. The intra-node scaling behaviour is analyzed by means of
weak-scaling experiments choosing the granular gas as a test problem and the
Emmy cluster as a test machine. The influence of the number of dimensions
in which the domain is partitioned is also only analyzed for this configuration.
All further scaling tests of the granular gas scenario use three-dimensional
domain partitionings. All inter-node weak-scaling experiments start with a single
node and extend to the full machine where possible. The experiments on the
SuperMUC supercomputer were obtained at the Extreme Scaling Workshop in
July~2013 at the LRZ, where at most 16~islands corresponding to 8\,192~nodes
were available. All strong-scaling experiments start on a single node except
on the Juqueen supercomputer, where we chose to start at 32~nodes which is
the minimum allocation unit in the batch system on Juqueen. The experiments
extend to a number of nodes where a notable efficiency degradation is
observed. Since the results on the SuperMUC were obtained well before the
other experiments, no scaling experiments with the hexagonal close packing
scenario were performed.

\subsection{Time-Step Profiles}
\label{sec:time-step_profiles}

\begin{figure}
	\includegraphics[width=.7\linewidth]{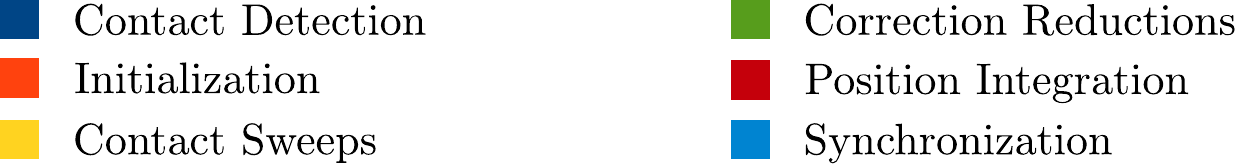}
	\centering
	\subfloat[Time-step profile of the granular gas executed with $5 \times 2 \times 2 = 20$ processes on a single node.]{%
		\includegraphics[width=.45\linewidth]{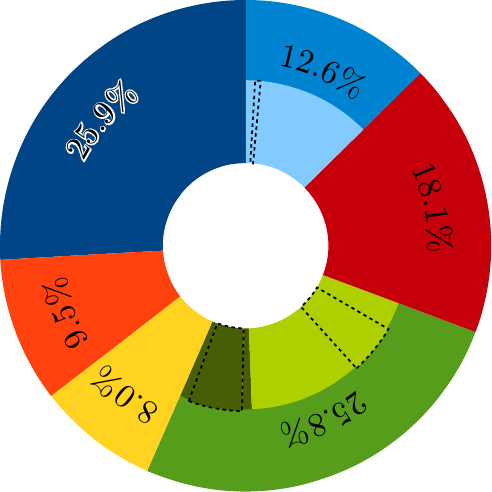}
		\label{fig:mpinano-timestep-components-1node}
	}\hfill
	\subfloat[Time-step profile of the granular gas executed with $8 \times 8 \times 5 = 320$ processes on 16 nodes.]{%
		\includegraphics[width=.45\linewidth]{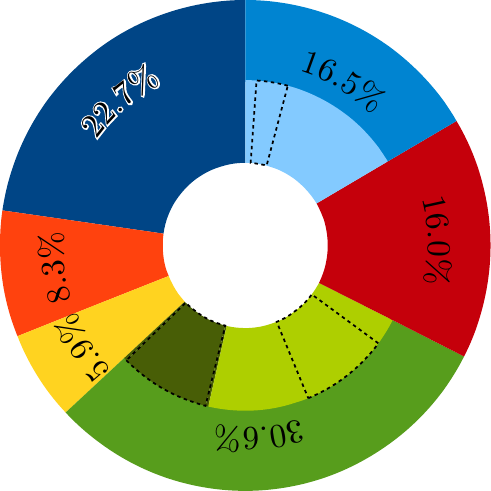}
		\label{fig:mpinano-timestep-components-16nodes}
	}
	\caption{The time-step profiles for two weak-scaling executions of the granular gas on the Emmy cluster with $25^3$ particles per process.}
	\label{fig:mpinano-timestep-components}
\end{figure}

In this section we clarify how much time is spent in the various phases of the
time-step procedure and how this time changes in a weak scaling depending on the
test problem. Fig.~\ref{fig:mpinano-timestep-components} breaks down the
wall-clock times of various time step components in two-level pie charts for the granular gas scenario. The
times are averaged over all time steps and processes.
The dark blue section corresponds to the fraction of the time in a time step
used for detecting and filtering contacts. The orange
section corresponds to the time used for initializing the velocity accumulators
and corrections. The time to relax the contacts is indicated by the yellow time slice. It includes
the contact sweeps for all 10~iterations without the correction reductions.
The time used by all correction reductions is shown in the green section which includes the
reductions for each iteration and the reduction after the initialization. The time slice is split up on the second level in
the time used for assembling, exchanging, and processing the first correction reduction
message (dark green section) and the time used for assembling, exchanging, and
processing the second correction reduction message (light green section). The time slices are
depicted counterclockwise in the given order. The message-exchange communications
have a dotted border to distinguish them from the rest. A single message-exchange
communication time measurement started, when sending the first message buffer to the neighbors,
and ended, when having received the last message buffer from the neighbors. The dark red section
corresponds to the time used by the time-integration of the positions, and the
final blue section indicates the time used by the position synchronization. The
latter is split up into assembling, exchanging, and processing of the message in
the inner ring. The message-exchange communication is highlighted by the dashed
border again. The first pie chart in Fig.~\ref{fig:mpinano-timestep-components-1node}
corresponds to the time-step profile of an execution in the weak-scaling
experiment with the three-dimensional domain partitioning $5 \times 2 \times 2$
on a single node of the Emmy cluster.
Fig.~\ref{fig:mpinano-timestep-components-16nodes} shows the time-step profile of
an execution in the weak-scaling experiment with the three-dimensional domain partitioning
$8 \times 8 \times 5$ on 16~nodes. The two time slices involving communication need
more time in comparison to Fig.~\ref{fig:mpinano-timestep-components-1node},
especially the framed slices on the second level which amount to the
communication. The wall-clock time for the components involving no communication
was roughly the same in both runs. The enlarged synchronization time-slices
in Fig.~\ref{fig:mpinano-timestep-components-16nodes} then approximately
amount to the increased time-step duration on 16 nodes. Overall, computations in the time
step of this granular gas scenario prevail. But since
the collision frequency is low, the 10~contact sweeps, marked by the yellow and green sections,
are dominated by communication.

\begin{figure}
	\includegraphics[width=.7\linewidth]{time-step-profile-legend}
	\centering
	\subfloat[Time-step profile of the hexagonal close packing scenario executed with $5 \times 2 \times 2 = 20$ processes on a single node.]{%
		\includegraphics[width=.45\linewidth]{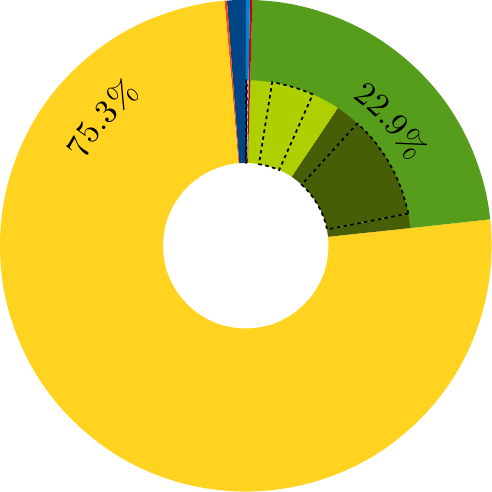}
		\label{fig:mpilattice-timestep-components-1node}
	}\hfill
	\subfloat[Time-step profile of the hexagonal close packing scenario executed with $8 \times 8 \times 5 = 320$ processes on 16 nodes.]{%
		\includegraphics[width=.45\linewidth]{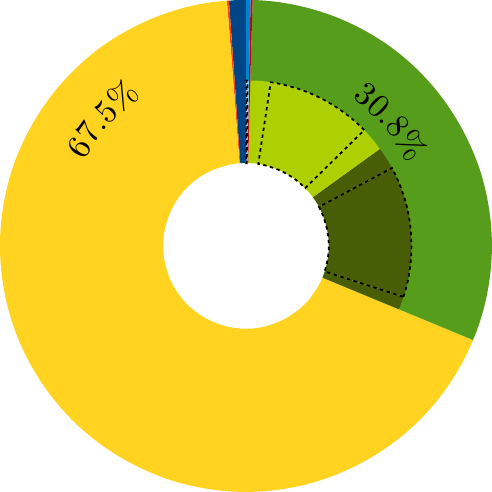}
		\label{fig:mpilattice-timestep-components-16nodes}
	}
	\caption{The time-step profiles for two weak-scaling executions of the hexagonal close packing scenario on the Emmy cluster with $10^3$ particles per process.}
	\label{fig:mpilattice-timestep-components}
\end{figure}

Fig.~\ref{fig:mpilattice-timestep-components} presents time-step
profiles for two weak-scaling executions of the hexagonal close packing scenario.
The time-step profiles use the same color coding as in Fig.~\ref{fig:mpinano-timestep-components}.
In contrast to the time-step profiles of the granular gas scenario, the time
step is dominated by the 100~contact sweeps (yellow section) and the
100~correction reductions (green section). Contact detection, position
integration, and synchronization play a negligible role. In
Fig.~\ref{fig:mpilattice-timestep-components-1node} the time-step profile of a
weak-scaling execution with again 20~processes on a single
node of the Emmy cluster is presented, whereas in
Fig.~\ref{fig:mpilattice-timestep-components-16nodes} the time-step profile of a
weak-scaling execution with again 320~processes on 16~nodes
is shown. The wall-clock time spent in the contact sweep was roughly the
same in both executions, hence the increased communication costs are mainly
responsible for the larger time slice of the correction reduction.

The time-step profiles showed that for the dilute granular gas scenario the
time spent in the various time-step components is well balanced and the time
spent in the communication routines moderately increases as the problem
size is increased. For the hexagonal close packings most of the time is spent
in the contact sweeps and the reduction of the velocity corrections. Components
such as the position integration and the final synchronization play a negligible
role due to the higher number of iterations in comparison to the granular gas
scenario.

\subsection{Weak-Scaling Results}
\label{sec:weak-scaling_results}

In the following subsections the weak-scaling results for both test problems on
the clusters are presented. Tab.~\ref{tab:weakscalingsummary} gives an overview
of the employed parameters.
\begin{table*}
\centering
\begin{tabular}{llllll}
	\toprule
	                                 & \multicolumn{3}{l}{Granular Gas}                          & \multicolumn{2}{l}{Hexagonal Close Packing} \\
	                                   \cmidrule(lr){2-4}                                          \cmidrule(lr){5-6}
	                                 & Emmy             & Juqueen          & SuperMUC            & Emmy             & Juqueen                  \\
	\midrule
	number of particles per process  & $25^3$           & $10^3$           & $10^3$              & $10^3$           & $10^3$                   \\
	number of time steps             & 1\,000           & 1\,000           & 10\,000             & 1\,000           & 100                      \\
	maximum number of particles      & $1.6 \cdot 10^8$ & $1.8 \cdot 10^9$ & $1.3 \cdot 10^8$    & $1.0 \cdot 10^7$ & $1.8 \cdot 10^9$         \\
	initial number of contacts       & 0                & 0                & 0                   & $6.0 \cdot 10^7$ & $1.1 \cdot 10^{10}$      \\
	solid volume fraction            & 23\%             & 23\%             & 3.8\%               & 74\%             & 74\%                     \\
	\bottomrule
\end{tabular}
\caption{Summary of the test problem parameters used for the weak-scaling experiments.\label{tab:weakscalingsummary}}
\end{table*}
The experiments differ in terms of the number of
particles generated per process depending on the amount of memory available.
In order to control the overall wall-clock time the number of
time steps performed varies between 100 and 10\,000. All wall-clock times
presented in the following subsections correspond to the average wall-clock
time needed to perform a single time step per 1\,000 particles facilitating
the comparison of the charts. The wall-clock times exclude the
time needed to setup the systems and generate the simulation output.
The scaling experiments of the granular gas
scenario on SuperMUC differs from the other granular gas experiments in that
the gas is considerably more dilute and a longer period of time is simulated.

\subsubsection{Granular Gas}

\begin{figure}
	\centering
	\includegraphics[width=\linewidth]{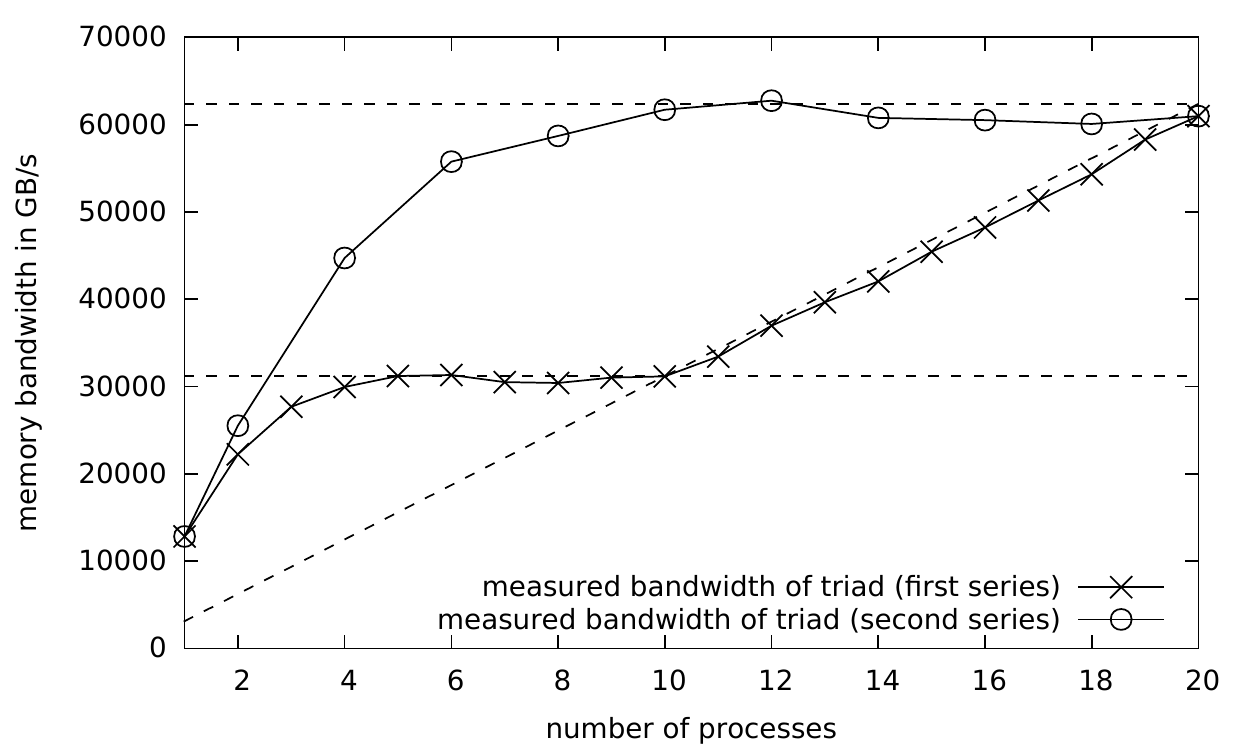}
	\caption{Measured bandwidth of the triad in the stream benchmark computed with a varying number of cores on a single node of the Emmy cluster.}
	\label{fig:stream-triad-emmy}
\end{figure}

First, we pay special attention to the intra-node weak-scaling
before turning to the inter-node weak-scaling since
the former is subject to the non-linear scaling behaviour of the memory bandwidth.
As a test problem we chose the granular gas scenario and as the test machine the
Emmy cluster. A single node in the cluster is equipped with
two processors each one having a single on-chip memory controller. The total
memory bandwidth available to both sockets is exactly twice the bandwidth of
a single socket. However, for a single socket a simple stream benchmark~\cite{mccalpin95}
reveals that the memory architecture is designed such that for $x$ cores
more than $\frac{1}{x}$ of the socket's total memory bandwidth is available.
Fig.~\ref{fig:stream-triad-emmy} plots the measured memory bandwidth of computations
of the triad as defined in the stream benchmark.
The computations were
performed by a varying number of cores in parallel. The first series of measurements
minimized the number of sockets in use meaning that the processor affinities
of the processes were adjusted such that all processes in measurements with less
or equal to 10~processes shared the same memory controller. In the second series
of measurements the processes were pinned to the sockets alternately such that
$2x$~processes had twice the bandwidth at their disposal as $x$ processes in the
first series. Indeed, the lower-left part of the first series' graph very well
matches the second series' graph with proper scaling. The measured bandwidth in
the first series increases for an increasing number of processes until the
available memory bandwidth of the first memory controller is saturated.
Measurements with more than 10~processes start to make use of the second memory
controller and the measured bandwidth continues to increase linearly.

\begin{figure}
	\centering
	\includegraphics[width=\linewidth]{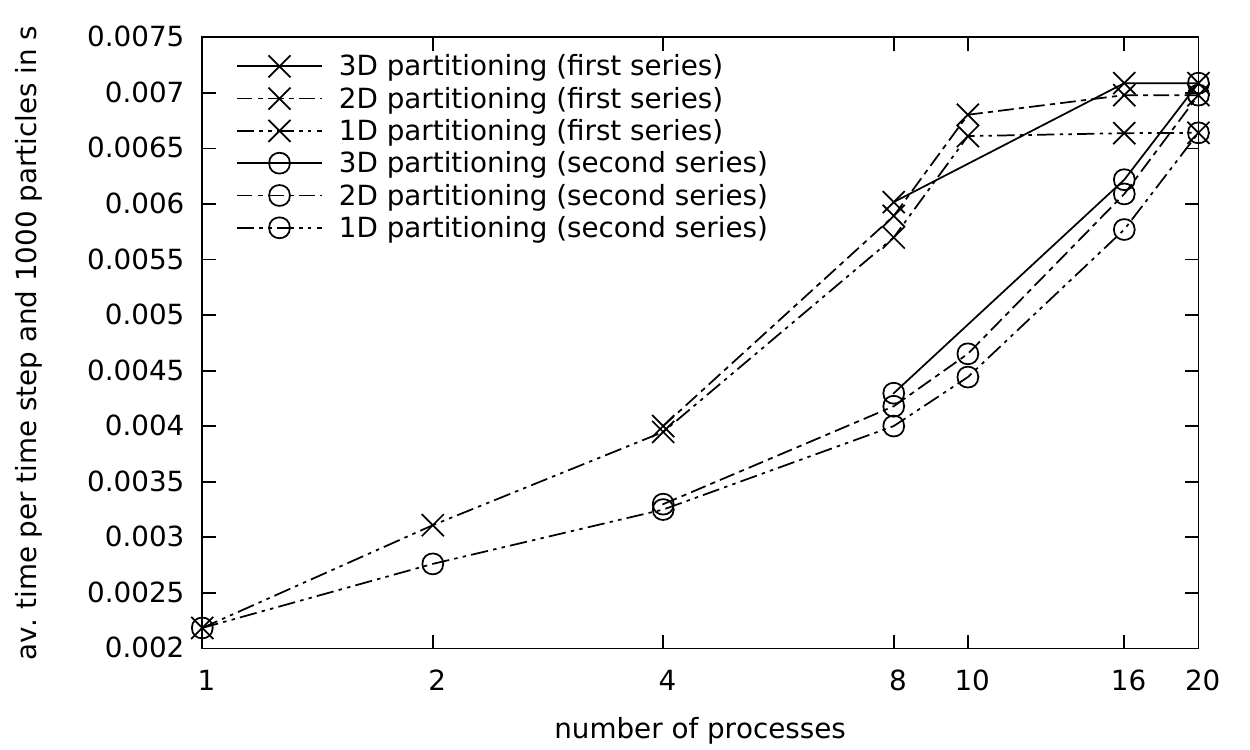}
	\caption{Intra-node weak-scaling graphs for a granular gas on the Emmy cluster.}
	\label{fig:mpinano-intranode-emmy}
\end{figure}

An analogous behavior can be observed in the intra-node weak-scaling graphs.
Fig.~\ref{fig:mpinano-intranode-emmy} plots the average wall-clock time needed
for a single time step and $1\,000$ particles.
In the first series of executions again the pinning strategy minimizing the
number of sockets in use was employed. The average wall-clock time needed per
time step increases considerably for executions with up to 10~processes. However,
beyond that point the weak-scaling graph continues almost ideally.
The second series of executions used as before the pinning strategy minimizing
the maximum number of processes per socket meaning that executions with $2x$
processes in the second series have twice the bandwidth at their disposal as the
executions with $x$ processes in the first series. The graphs of the second
series show that the wall-clock times needed per time step on $2x$ processes
indeed closely match the wall-clock times on $x$ processes in the first series.
This indicates that our implementation is limited by the available memory bandwidth.

The figure also distinguishes between weak-scaling graphs with one-, two-, and
three-dimensional domain partitionings since their communication volumes differ.
Higher-dimensional non-periodic domain partitionings have typically a higher
communication volume in comparison to lower dimensional non-periodic domain
partitionings with the same number of processes, due to the larger area of the
interfaces between the subdomains. The plotted timings for the one-dimensional domain
partitionings are indeed consistently slightly better than the timings for
two-dimensional domain partitionings, which are in turn slightly better than the
timings for three-dimensional domain partitionings.

Even though the intra-node weak-scaling results reveal an underperforming parallel efficiency
between 30.8\% and 32.9\% when computing on all cores of an Emmy node, the
correlation with the measured memory bandwidth of a triad suggests that a good
intra-node scaling can be expected as long as the available bandwidth scales. With
corresponding pinning this is the case as off the first full socket on the Emmy cluster.

\begin{figure}
	\centering
	\subfloat[Weak-scaling graph on the Emmy cluster.]{%
		\includegraphics[width=\linewidth]{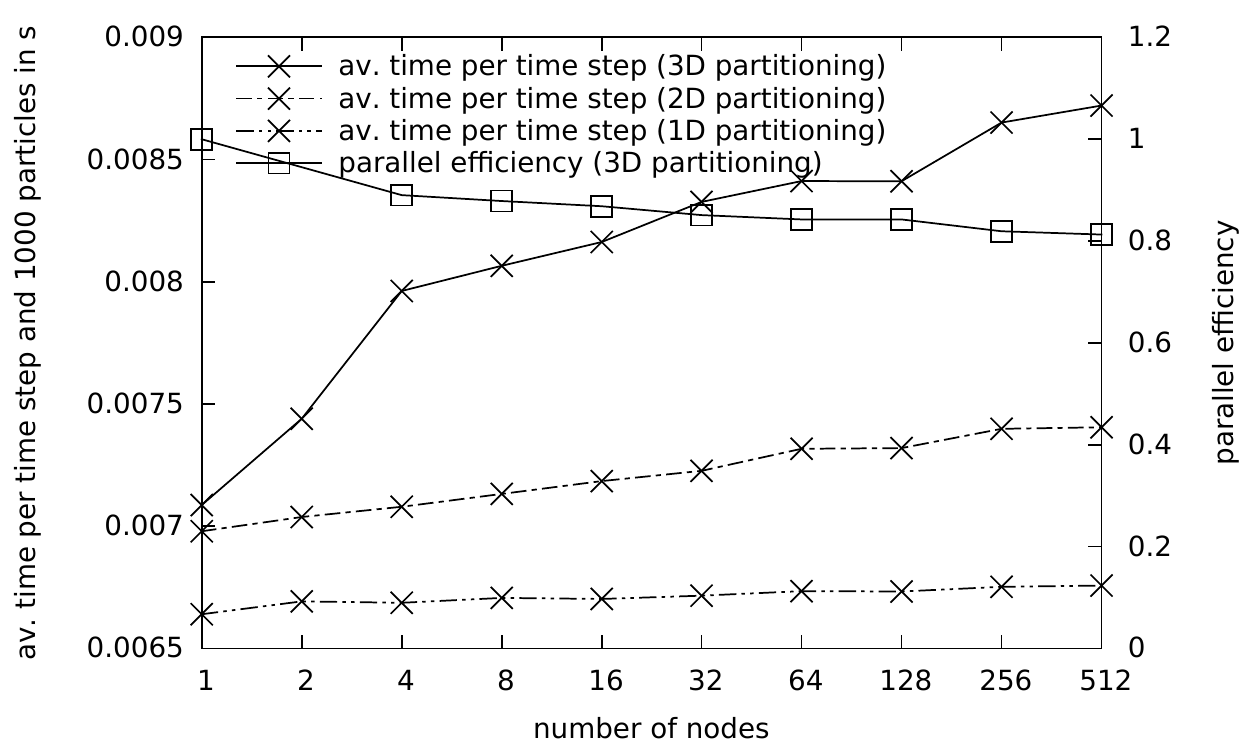}
		\label{fig:mpinano-internode-emmy}
	}\hfill
	\subfloat[Weak-scaling graph on the Juqueen supercomputer.]{%
		\includegraphics[width=\linewidth]{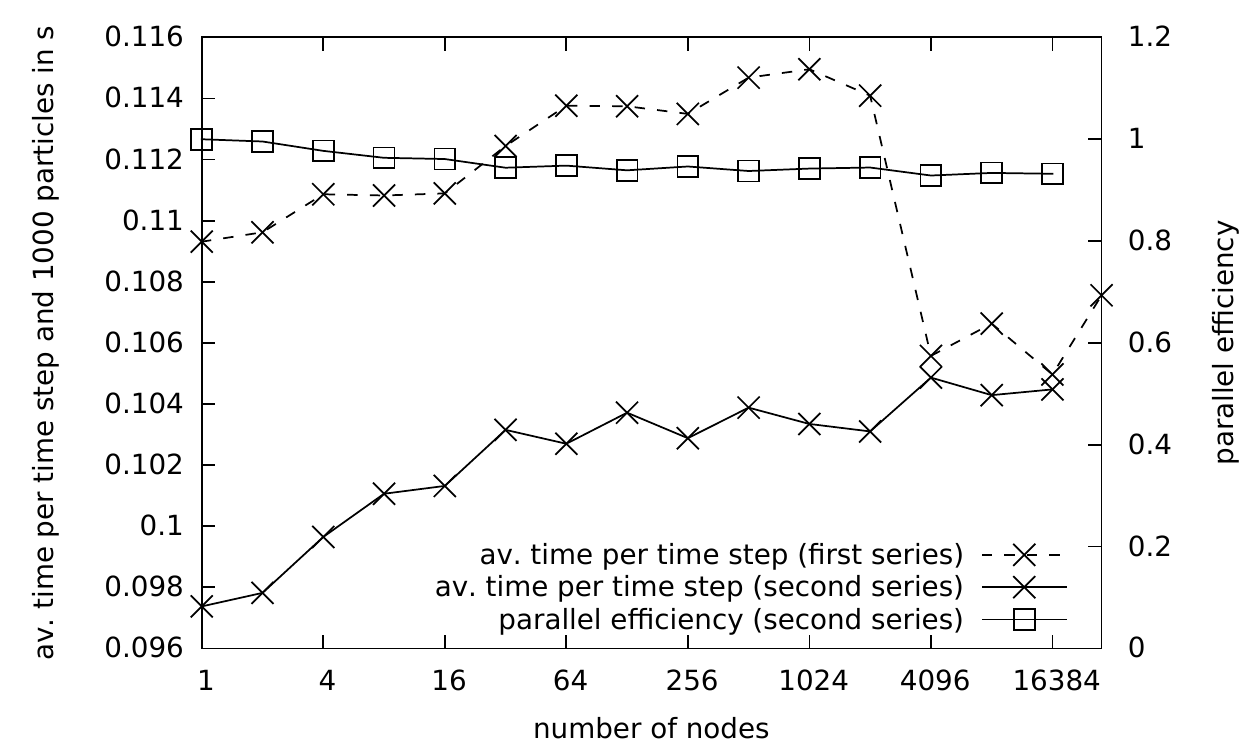}
		\label{fig:mpinano-internode-juqueen}
	}\hfill
	\subfloat[Weak-scaling graph on the SuperMUC supercomputer.]{%
		\includegraphics[width=\linewidth]{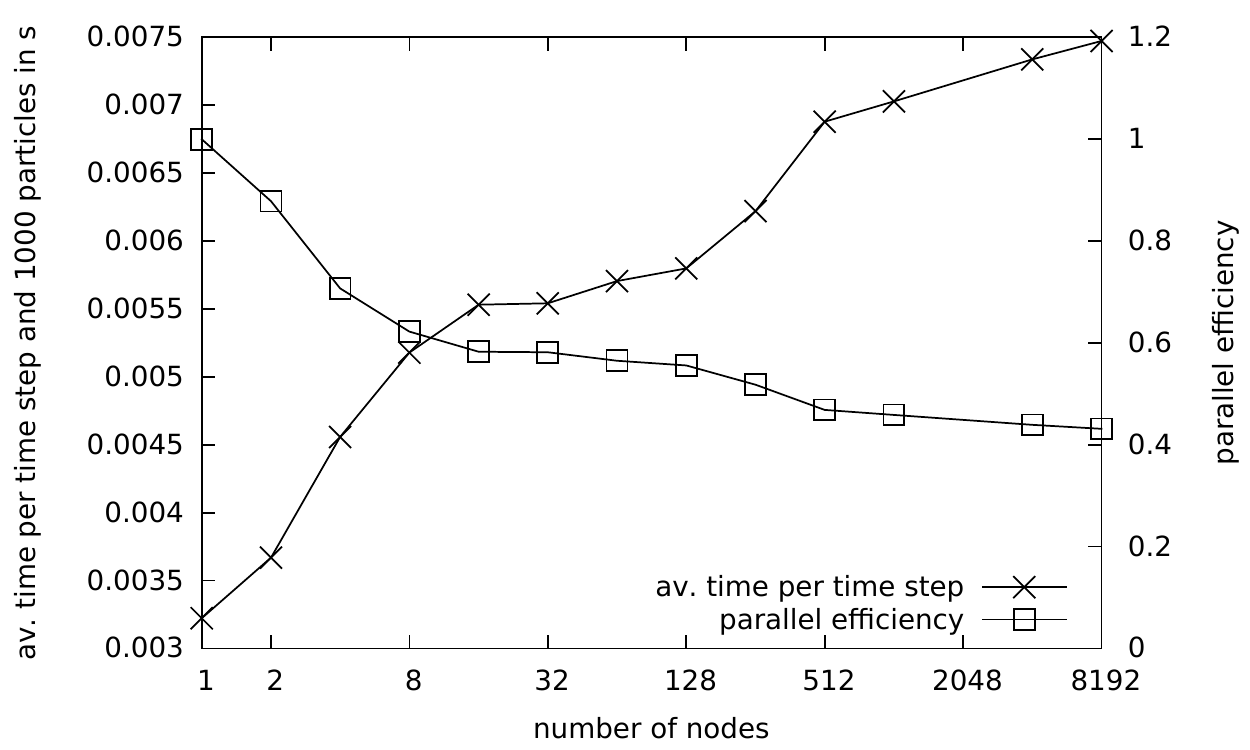}
		\label{fig:mpinano-internode-supermuc}
	}
	\caption{Inter-node weak-scaling graphs for a granular gas on all test machines.}
\end{figure}

Fig.~\ref{fig:mpinano-internode-emmy} extends the weak-scaling experiment to almost
the full Emmy cluster for one-, two-, and three-dimensional domain partitionings.
The scaling experiment for the one-dimensional domain partitionings performs best and
achieves on 512~nodes a parallel efficiency of 98.3\% with respect to the
single-node performance. The time measurements for two-dimensional domain
partitionings are consistently slower, but the parallel efficiency does not drop
below 89.7\%. The time measurements for three-dimensional domain partitionings
come in last, and the parallel efficiency goes down to 76.1\% for 512~nodes.
This behaviour can be explained by the differences in the communication
volumes of one-, two-, and three-dimensional domain partitionings.
The results attest that the problem can be efficiently scaled (almost) up to the full
machine if the load per process is sufficiently large.

Fig.~\ref{fig:mpinano-internode-juqueen} shows the results of the inter-node
weak-scaling experiments on the Juqueen supercomputer.
The scaling experiments were only performed with the more demanding
three-dimensional domain parititionings. In the first
series of measurements the average wall-clock time per time step increases as
expected up to 2\,048~nodes. But then the average time-step duration for setups
with 4\,096~nodes and beyond is significantly shorter than the average time-step
duration with fewer
nodes. The time steps are even computed faster than on a single node, where no
inter-node communication takes place at all. Assuming that intra-node
communication is faster than inter-node communication, this is a puzzling
result. In fact, it turned out that the intra-node communication was
responsible for the behaviour: The default mechanism for intra-node
communication is via shared memory on the Juqueen. In the second series
of measurements we disallowed the usage of shared memory for intra-node communication.
This resulted in the measurements that are consistently faster than the
measurements from the first series, and the parallel efficiency is
more or less monotonically decreasing with an excellent parallel efficiency of at least 92.9\%.

The reason why the measured times in the first
series became shorter for 4\,096~nodes and more is revealed when considering how
the processes get
mapped to the hardware. The default mapping on Juqueen is ABCDET, where the
letters A to E stand for the five dimensions of the torus network, and T stands
for the hardware thread within each node. The six-dimensional coordinates are
then mapped to the MPI ranks in a row-major order, that is, the last dimension
increases fastest. The T coordinate is limited by the number of processes per
node, which was 64 for the above measurements. Upon creation of a three-dimensional
communicator, the three dimensions of the domain partitioning are mapped also in
row-major order. This effects, if the number of processes in z-dimension
is less than the number of processes per node, that a two-dimensional or
even three-dimensional section of the domain partitioning is mapped to a single
node. However, if the number of processes in z-dimension is larger or equal to
the number of processes per node, only a one-dimensional section of the domain
partitioning is mapped to a single node. A one-dimensional section of the
domain partitioning performs considerably less intra-node communication than
a two- or three-dimensional section of the domain partitioning. This matches
exactly the situation for 2\,048 and 4\,096 nodes. For 2\,048~nodes, a
two-dimensional section $1 \times 2 \times 32$ of the domain partitioning
$64 \times 64 \times 32$ is mapped to each node, and for 4\,096~nodes a
one-dimensional section $1 \times 1 \times 64$ of the domain partitioning
$64 \times 64 \times 64$ is mapped to each node. To substantiate this
claim, we confirmed that the performance jump occurs when the last dimension of the
domain partitioning reaches the number of processes per node, also when using
16 and 32~processes per node.

Fig.~\ref{fig:mpinano-internode-supermuc} presents the weak-scaling results on
the SuperMUC supercomputer. The setup differs from the granular gas scenario
presented in Sect.~\ref{sec:granular_gas} in that it is more dilute. The
distance between the centers of two granular particles along each spatial
dimension is~$\unit[2]{\centi\meter}$, amounting to a solid volume fraction of
3.8\% and consequently to less collisions. As on the Juqueen supercomputer
only three-dimensional domain partitionings were used.
All runs on up to 512~nodes were running within a single island. The run
on 1\,024~nodes also used the minimum number of 2~islands. The run on
4\,096~nodes used nodes from 9~islands, and the run on 8\,192~nodes used nodes
from 17~islands, that is both runs used one island more than required.
The graph shows that most of the performance is lost in runs on
up to 512~nodes. In these runs only the non-blocking intra-island communication
is utilised. Thus this part of the setup is very similar to the Emmy
cluster since it also has dual-socket nodes with Intel Xeon E5 processors and a
non-blocking tree Infiniband network. Nevertheless, the intra-island scaling
results are distinctly worse. The reasons for these differences were not yet
further investigated. However, the scaling behaviour beyond a single island
is decent featuring a parallel efficiency of 73.8\% with respect to a single
island. A possible explanation of the underperforming intra-node scaling behaviour could be
that some of the Infiniband links were degraded to QDR, which was a known
problem at the time the extreme-scaling workshop took place. The communication
routines then need $\frac{5 \cdot 64}{4 \cdot 66} \approx 1.21$ times longer to
complete. This could also explain the high variability of the runs' wall-clock
times.

Subsequently, a second series of measurements was performed with $60^3$
non-spherical particles per process. The scaling behaviour is comparable
to the scaling behaviour observed in Fig.~\ref{fig:mpinano-internode-supermuc}.
However, the largest weak-scaling run simulated
$28\,311\,552\,000 \approx 2.8 \cdot 10^{10}$ non-spherical particles --
a possibly record-breaking number for non-smooth contact dynamics.

\subsubsection{Hexagonal Close Packings of Spheres}

\begin{figure}
	\centering
	\subfloat[Weak-scaling graph on the Emmy cluster.]{%
		\includegraphics[width=\linewidth]{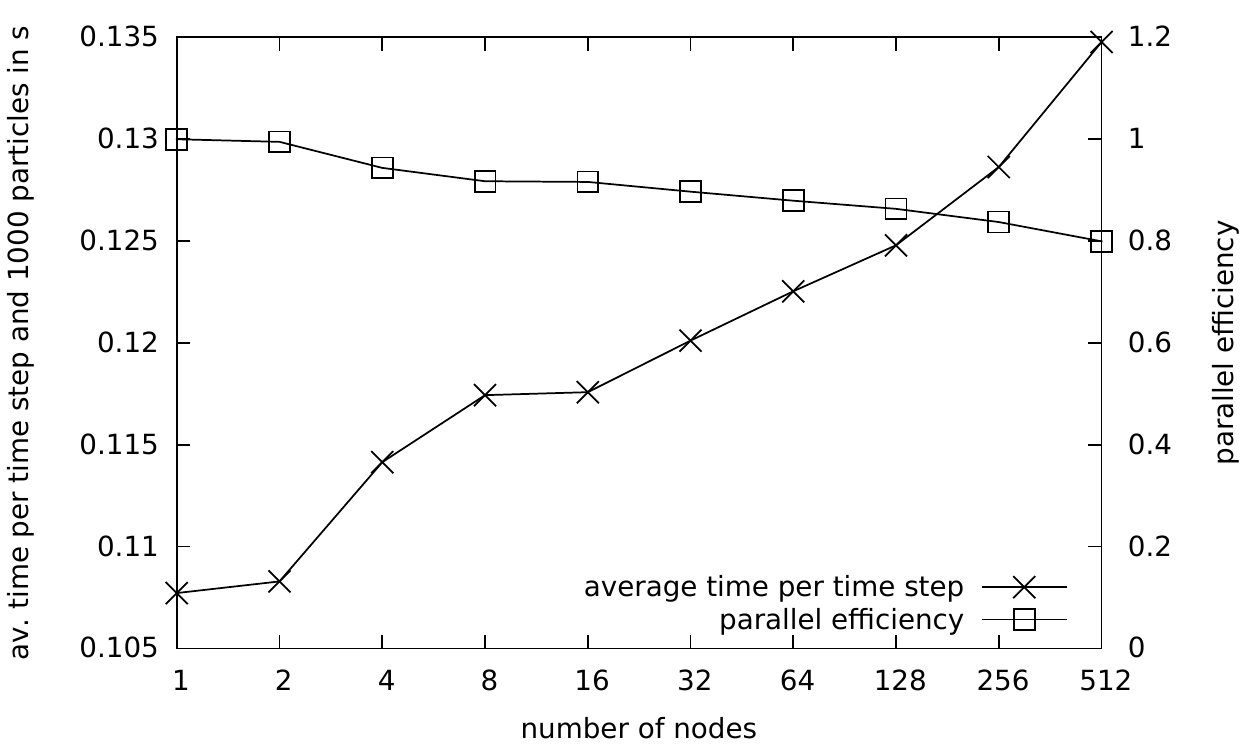}
		\label{fig:mpilattice-internode-emmy}
	}\hfill
	\subfloat[Weak-scaling graph on the Juqueen supercomputer.]{%
		\includegraphics[width=\linewidth]{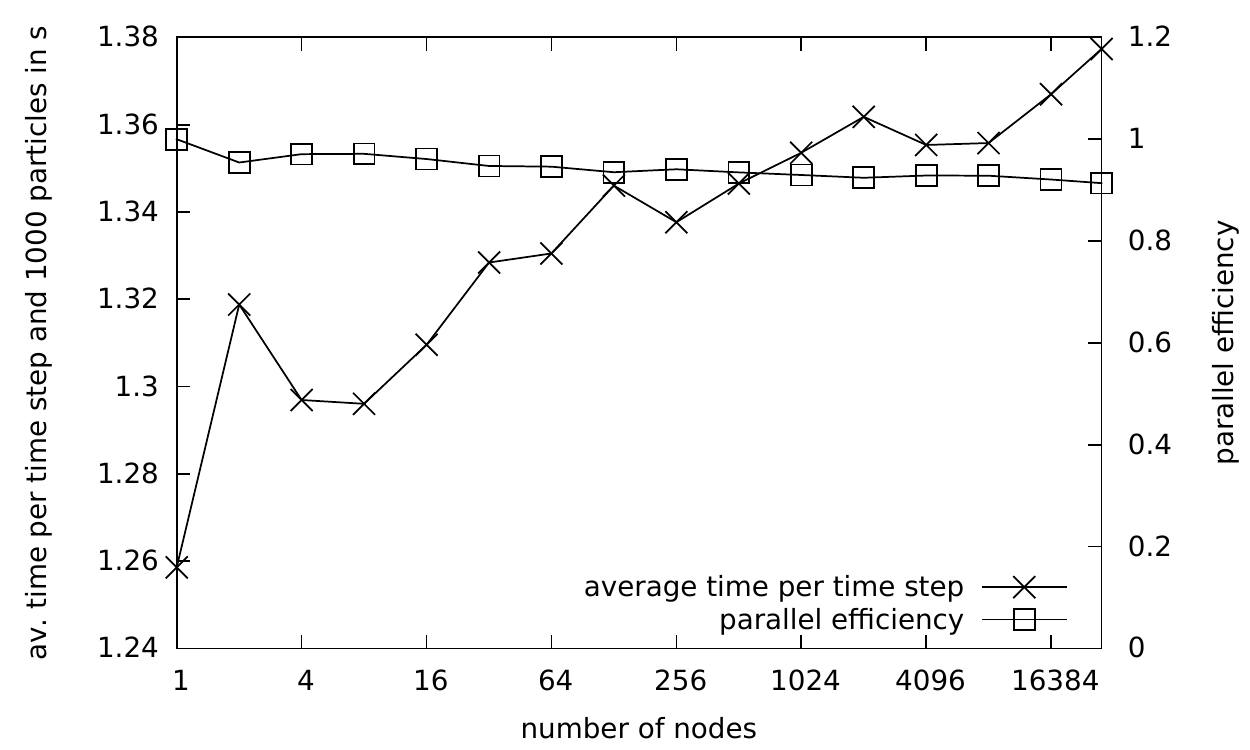}
		\label{fig:mpilattice-internode-juqueen}
	}
	\caption{Inter-node weak-scaling graphs for hexagonal close packings of spheres.}
\end{figure}

Fig.~\ref{fig:mpilattice-internode-emmy} shows the average wall-clock time
needed for a single time step in the hexagonal close packing test on the Emmy cluster.
The parallel efficiency with respect to a single node remains above 79.9\% for
all executions. This is slightly better than the parallel efficiency
of 76.1\% for the granular gas.

The weak-scaling results of the hexagonal close packing scenario on
the Juqueen supercomputer are presented in Fig.~\ref{fig:mpilattice-internode-juqueen}.
The parallel efficiency with respect to a single node stayed above
91.4\% for all measurements. This result is almost as good as the 92.9\%
parallel efficiency in the scaling experiments of the granular gas.
The largest execution ran $1\,024 \times 1\,792 \times 1 = 1\,835\,008$~processes
on all 28\,672~nodes of the machine, where $10\,240 \times 17\,920 \times 10 = 1\,835\,008\,000$~particles
were spawned, in total leading to $10\,826\,547\,200 \approx 1.1 \cdot 10^{10}$
contacts -- again a possibly record-breaking number for non-smooth contact dynamics.

\subsection{Strong-Scaling Results}
\label{sec:strong-scaling_results}

In the following subsections the strong-scaling results for both test problems on
the clusters are presented. Tab.~\ref{tab:strongscalingsummary} gives an overview
of the employed parameters.
\begin{table*}
\centering
\begin{tabular}{llllll}
	\toprule
	                                 & \multicolumn{3}{l}{Granular Gas}                          & \multicolumn{2}{l}{Hexagonal Close Packing} \\
	                                   \cmidrule(lr){2-4}                                          \cmidrule(lr){5-6}
	                                 & Emmy                        & Juqueen                     & SuperMUC                    & Emmy                          & Juqueen                  \\
	\midrule
	number of particles              & $320 \times 160 \times 160$ & $320 \times 320 \times 320$ & $128 \times 128 \times 128$ & $1\,280 \times 640 \times 10$ & $2\,048 \times 2\,048 \times 10$ \\
	number of time steps             & 1\,000                      & 1\,000                      & 100                         & 50                            & 20               \\
	solid volume fraction            & 23\%                        & 23\%                        & 3.8\%                       & 74\%                          & 74\%             \\
	\bottomrule
\end{tabular}
\caption{Summary of the test problem parameters used for the strong-scaling experiments.\label{tab:strongscalingsummary}}
\end{table*}
The experiments differ in terms of the number of
particles generated in total and the number of time steps used for averaging.
As in the weak-scaling experiments the granular gas scenario on SuperMUC
is considerably more dilute than on the other machines.

\subsubsection{Granular Gas}

\begin{figure}
	\centering
	\subfloat[Strong-scaling graph on the Emmy cluster.]{%
		\includegraphics[width=\linewidth]{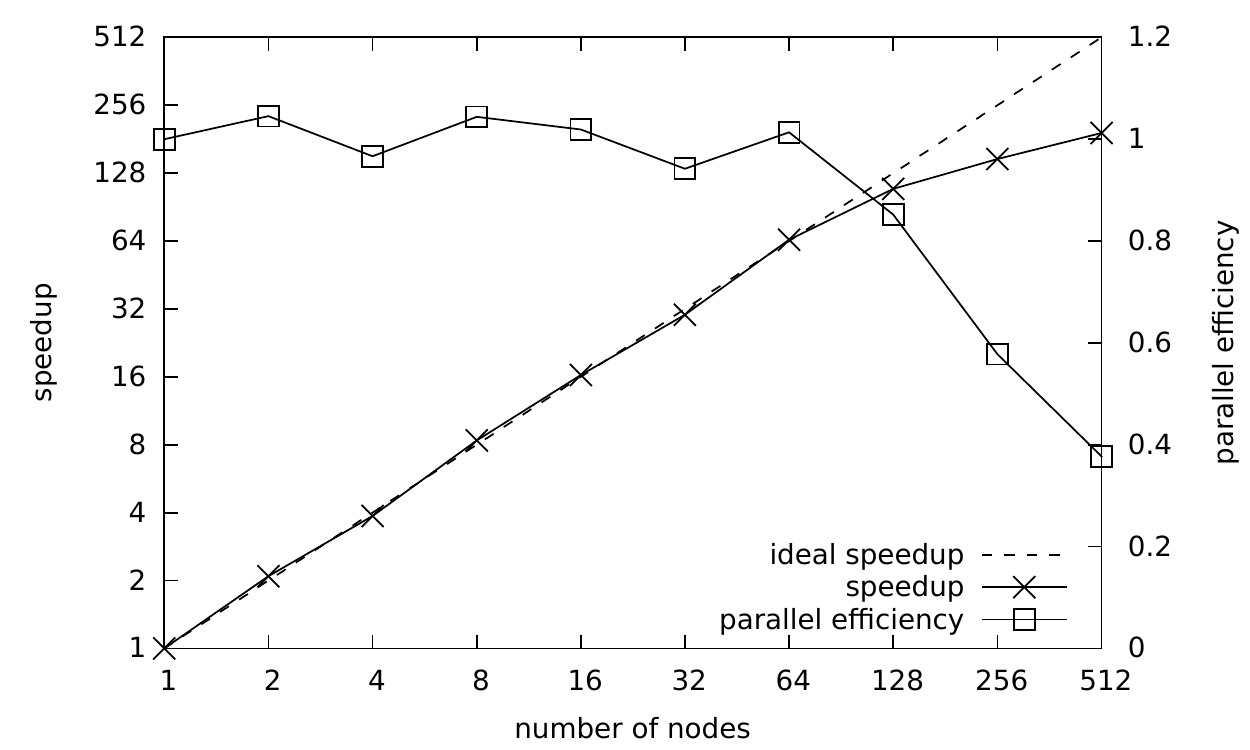}
		\label{fig:mpinano-strong-emmy}
	}\hfill
	\subfloat[Strong-scaling graph on the Juqueen supercomputer.]{%
		\includegraphics[width=\linewidth]{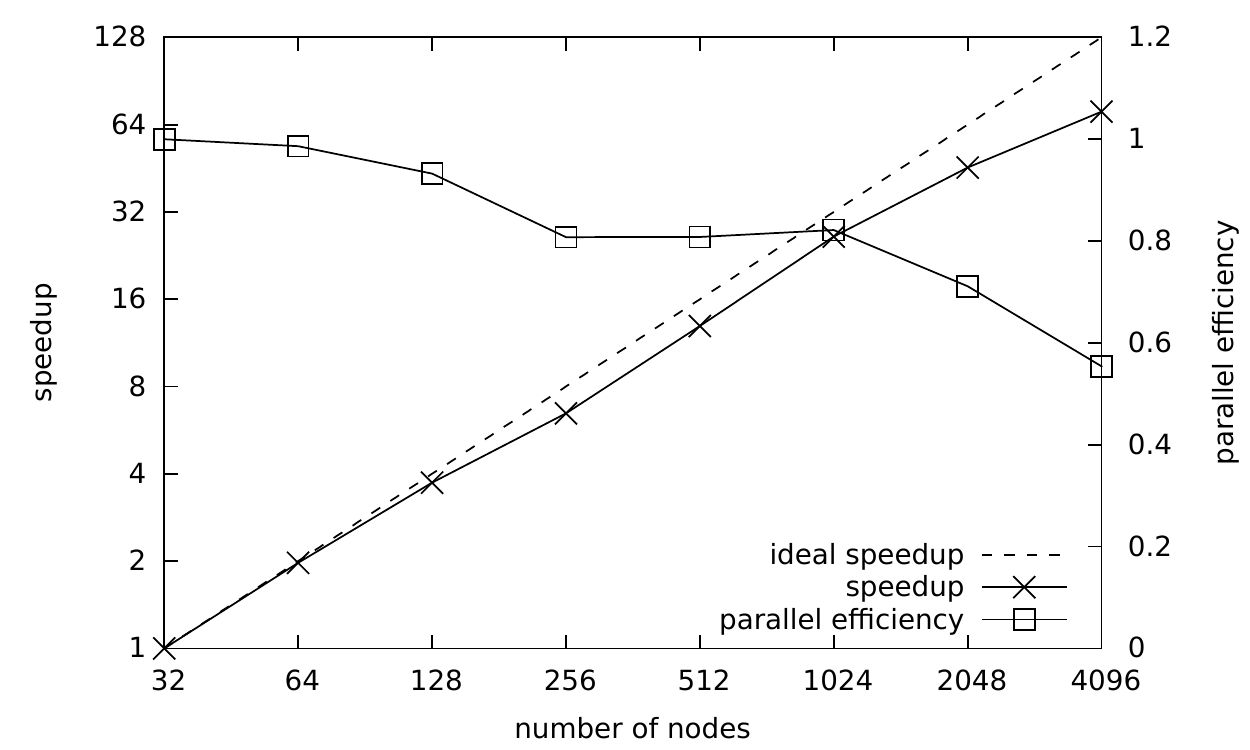}
		\label{fig:mpinano-strong-juqueen}
	}\hfill
	\subfloat[Strong-scaling graph on the SuperMUC supercomputer.]{%
		\includegraphics[width=\linewidth]{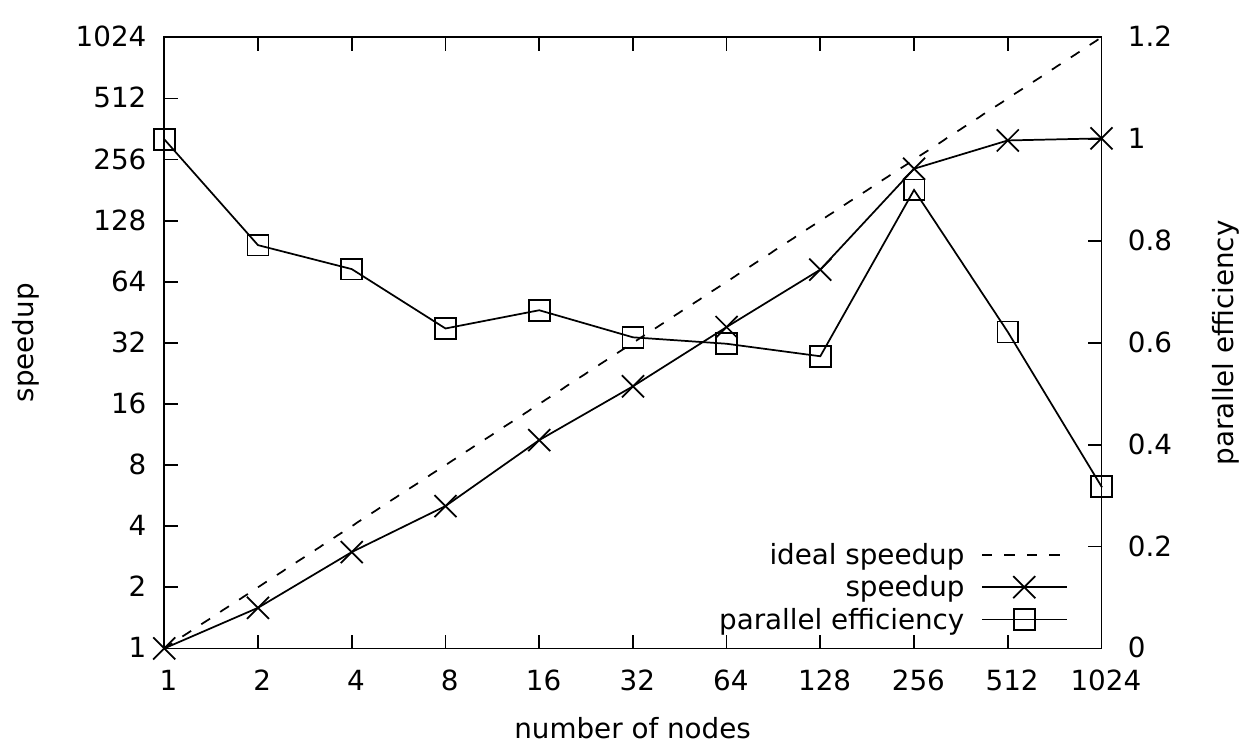}
		\label{fig:mpinano-strong-supermuc}
	}
	\caption{Strong-scaling graphs for a granular gas test problem on all test machines.\label{fig:mpinano-strong}}
\end{figure}

Fig.~\ref{fig:mpinano-strong} presents the strong-scaling results of the
granular gas scenario on all clusters.
The strong-scaling graph on the
Emmy cluster is presented in Fig.~\ref{fig:mpinano-strong-emmy}.
A total of $320 \times 160 \times 160 = 8\,192\,000$
particles was used, leading to at most $64 \times 80 \times 80 = 409\,600$ particles
per process on a single node and at least $10 \times 8 \times 10 = 800$ particles
per process on 512~nodes. The speedup is ideal for up to 64~nodes and then
gradually becomes more inefficient. However, no turnover is observed.
Some time measurements exceed the optimal speedup, which can happen for example
if the problem becomes small enough to fit into one of the caches. In conclusion,
the scaling experiments for this dilute setup on the Emmy cluster
suggest that one obtains a satisfactory parallel efficiency on the whole cluster,
as long as several thousand particles are present per process.

Fig.~\ref{fig:mpinano-strong-juqueen} presents the results of the strong-scaling
experiments on the Juqueen supercomputer for the granular gas. The total number
of particles was 32\,768\,000 particles. In the execution on 32~nodes each of
the $16 \times 16 \times 8 = 2\,048$~processes
initially had $20 \times 20 \times 40 = 16\,000$ non-spherical particles, and
in the execution on 4\,096~nodes each of the $64 \times 64 \times 64 = 262\,144$~processes
spawned $5 \times 5 \times 5 = 125$~particles. The parallel efficiency is
plotted with respect to 32~nodes and stays above 80.7\% for up to 1\,024~nodes and 500~particles per process
before rapidly decreasing. On 4\,096~nodes the efficiency is at 55.4\%.
The weak- and strong-scaling results are both better in comparison to
the Emmy cluster, owed to the torus network which performs excellent for
the nearest-neighbor communication.

The results of the strong-scaling experiments on the SuperMUC supercomputer are
shown in Fig.~\ref{fig:mpinano-strong-supermuc}.
In total $128^3$~non-spherical particles were
simulated. Hence, in the single-node run each process owned
$32 \times 64 \times 64 = 131\,072$ particles, and in the run on 1\,024~nodes,
each process owned $8 \times 4 \times 4 = 128$~particles. The parallel
efficiency is at 90.0\% on 256~nodes. Beyond that point it decreases
dramatically, indicating that the scaling is fine as long as at least about
500~particles are present per process.

\subsubsection{Hexagonal Close Packings of Spheres}

\begin{figure}
	\centering
	\subfloat[Strong-scaling graph on the Emmy cluster.]{%
		\includegraphics[width=\linewidth]{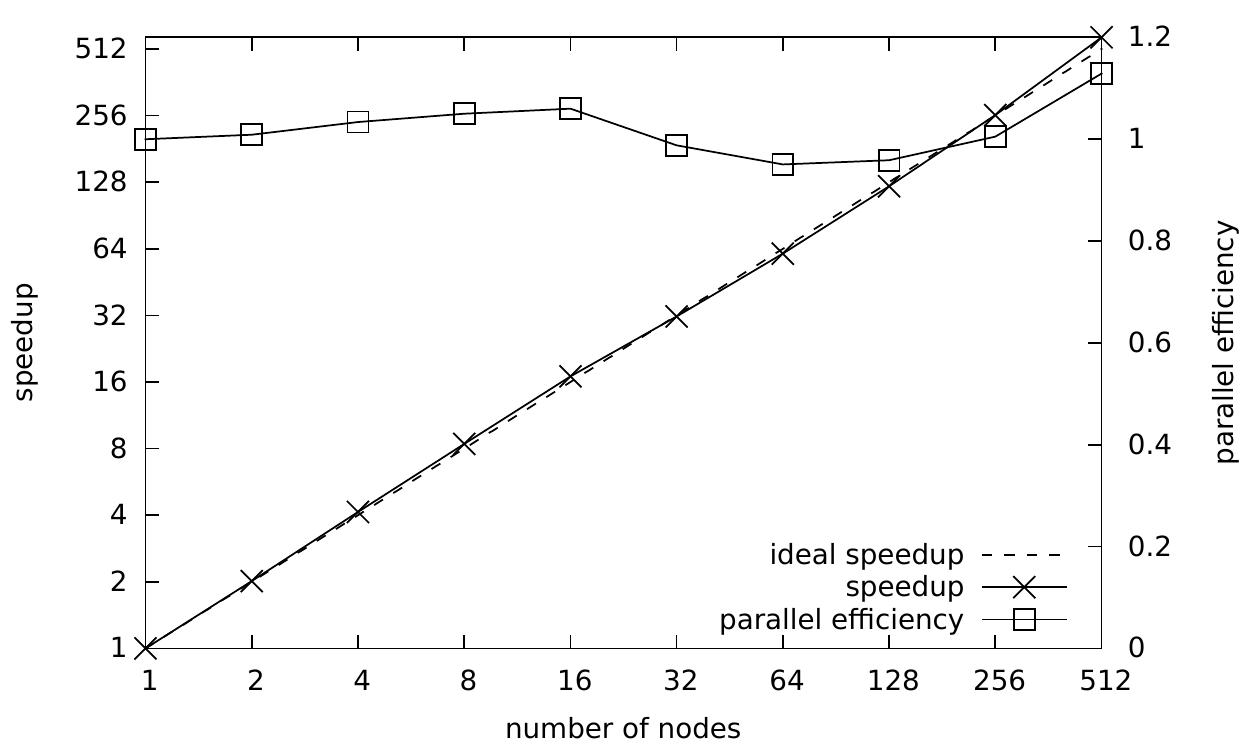}
		\label{fig:mpilattice-strong-emmy}
	}\hfill
	\subfloat[Strong-scaling graph on the Juqueen supercomputer.]{%
		\includegraphics[width=\linewidth]{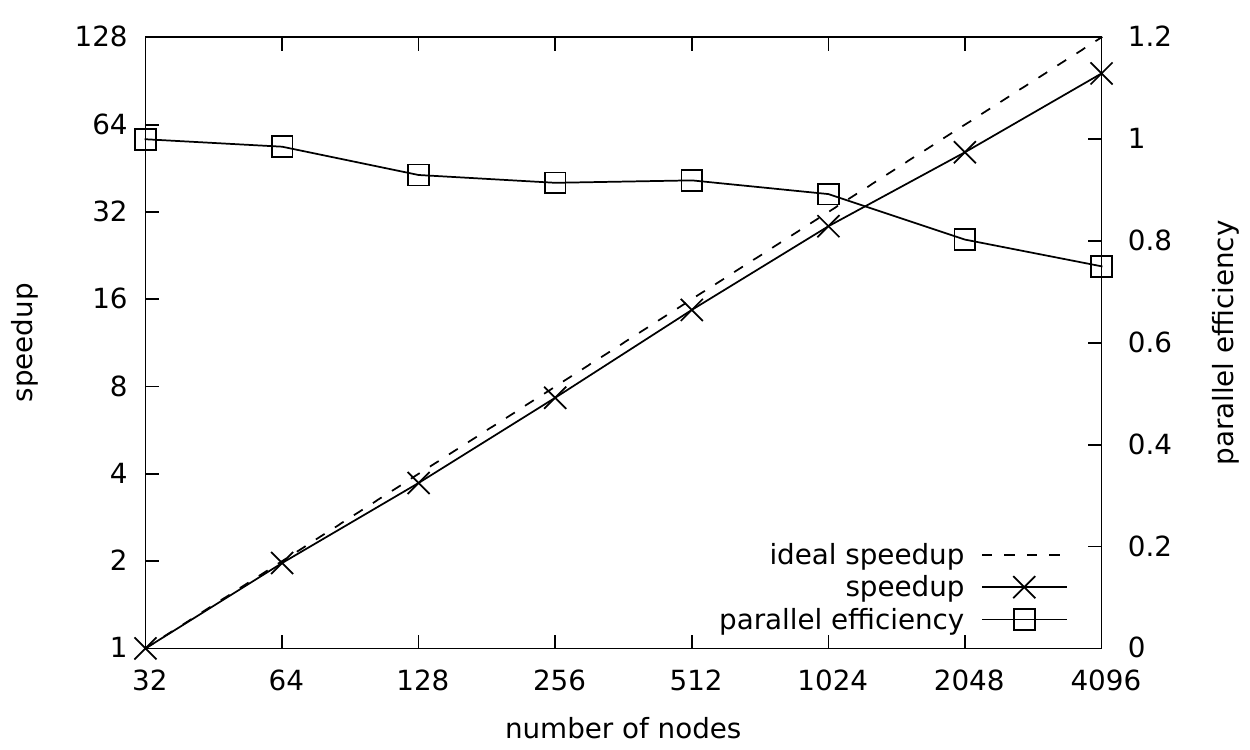}
		\label{fig:mpilattice-strong-juqueen}
	}
	\caption{Strong-scaling graphs for hexagonal close packings of spheres.}
\end{figure}

In the strong-scaling experiment on the Emmy cluster in total $1\,280 \times 640 \times 10 = 8\,192\,000$ particles
were generated. The experiment was run for 1 to 512~nodes, such that the
smallest setup with $5 \times 4 \times 1 = 20$~processes on a single node
generated $256 \times 160 \times 10 = 409\,600$ spherical particles per process,
and the largest setup with $128 \times 80 \times 1 = 10\,240$ processes on
512~nodes generated $10 \times 8 \times 10 = 800$ particles per process.
Fig.~\ref{fig:mpilattice-strong-emmy} presents the results.
A super-linear speedup is observed for several executions, which is likely
due to caching effects, since the working set size becomes very small.
In the strong-scaling experiment for 512~nodes of the granular gas scenario on Emmy
also only 800~particles were generated per process. However, the
computational intensity here is much higher in comparison to that of the granular gas, because far
more contacts have to be resolved. This explains the high parallel
efficiency of 113\% in comparison to the disappointing parallel efficiency of
37.7\% from Fig.~\ref{fig:mpinano-strong-emmy}. In conclusion, the scaling
experiments suggest that a few hundred particles per
process are enough to achieve a very good parallel efficiency on the Emmy cluster
if the granular flow is dense.

For the strong-scaling experiment on the Juqueen supercomputer a hexagonal close packing with 41\,943\,040~particles in total was
created. The smallest execution ran $64 \times 32 \times 1 = 2\,048$~processes on 32~nodes,
where $32 \times 64 \times 10 = 20\,480$ spherical particles were generated
per process. The largest execution ran $512 \times 512 \times 1 = 262\,144$~processes on 4\,096~nodes,
where $4 \times 4 \times 10 = 160$~particles were generated per process.
Fig.~\ref{fig:mpilattice-strong-juqueen} shows the speedup and the
parallel efficiency on the second axis, both with respect to 32~nodes.
A parallel efficiency of 75.0\% on 4\,096~nodes is achieved, where only
160~particles were owned per process. This suggests that a reasonable good
efficiency can be achieved for a dense setup on the Juqueen supercomputer, as
long as several hundred particles are created per process.

\section{Related Work}
\label{sec:related_work}

Other authors have proposed approaches for parallelizing non-smooth contact
dynamics on architectures with distributed memory. All of them are based
on domain partitionings. A parallelization strategy termed non-smooth contact
domain decomposition~(NSCDD) implemented in the renowned LMGC90 code was
lately presented in~\cite{visseq12,visseq13} by Visseq et al. The approach is
inspired by the finite element tearing and interconnect~(FETI) method for
solving partial differential equations in computational mechanics.
The authors suggest to decouple the multi-contact problem such that on each
process a multi-contact problem is solved having the same structure as a
multi-contact problem that is solved sequentially. Particles with multiple
contacts that are associated with different subdomains are duplicated, similar to
shadow copies %
used in this article.
However, the mass and inertia are split among all
instantiations.
The coupling is recovered by adding linear equations gluing the
duplicates back together through additional Lagrange multipliers.
In contrast
to the contact constraints, the interface equations are linear, and
a block-diagonal system of linear equations 
must be solved after several
sweeps over all contacts. 
In \cite{visseq13}, the authors present
simulations with up to $2\cdot10^5$ spherical particles and $2\cdot 10^6$
contacts, time-integrated on up to 100~processes.
The NSCDD allows non-nearest-neighbor communication in order to allow enlarged rigid bodies instead of
introducing a concept analogous to global bodies.

Prior to Visseq et al., Koziara et al.\ presented the parallelization
implemented in the solfec code~\cite{koziara11}. This approach dispenses with
the separation into interface problems and local multi-contact problems.
A classic NBGS is parallelized with a non-negligible but inevitable amount of serialization.
Bodies are instantiated redundantly on all processes, prohibiting scaling beyond
the memory limit. Instead of using accumulator and correction variables, as proposed in this
paper, the authors synchronize dummy 
particles (
particles that are in contact with
shadow copies or original instances) in addition to shadow copies in
order to implement contact shadow copies. As in the NSCDD, the system matrix (Delassus operator)
is set up explicitly instead of using matrix-free computations as proposed
here.
Simulations are presented with up to $1\cdot10^4$ polyhedral 
particles or
$6\cdot10^5$ contacts time-integrated on up to 64~processes.

At the same time, Shojaaee et al.\ presented another domain partitioning method
in~\cite{shojaaee12}. The presentation is restricted to two-dimensional
problems. The solver in the paper corresponds to a subdomain NBGS with relaxation
parameter $\omega = 1$, where the authors argue that divergence does typically
not occur. At least for three-dimensional simulations this is in our experience
not sufficient. Shadow copies are created not only if the hulls overlap the
neighboring subdomain but also if the particles approach the subdomain
boundaries, simplifying the intersection testing but introducing excessive shadow
copies. Shojaaee et al.\ also introduce contact shadow copies instead of using
accumulator and correction variables as proposed here.
Simulations
are presented with up to $1\cdot10^6$ circular particles in a dense packing on
up to 256~processes.

The approach presented in this paper improves in general the robustness and
scalability of previously published parallel algorithms. The
matrix-free approach facilitates the evaluation of the particle wrenches
in parallel as suggested in Sect.~\ref{sec:accumulator_and_correction_variables}
and thus reduces the amount of communicated data. The separation of bodies into
global and local bodies allows to restrict message-exchange communications to
nearest neighbors as detailed in Sect.~\ref{sec:nearest_neighbor_communication}
and thus maps well to various interconnect networks. Furthermore, the
synchronization protocol defined in Sect.~\ref{sec:shadow_copies} and
Sect.~\ref{sec:time-integration_and_synchronization} is not susceptible to numerical
errors in contrast to the conventional rules which are based on contact
locations. Last but not least the scaling experiments from Sect.~\ref{sec:scaling_experiments}
with up to $2.8\cdot10^{10}$ non-spherical particles or $1.1\cdot10^{10}$
contacts on up to $1.8\cdot10^{6}$ processes exceed all previously published
numbers by a factor of $10^3$ to $10^4$.


\section{Summary}
\label{sec:summary}

This article presents models and algorithms for performing scalable direct
numerical simulations of granular matter in hard contact as we implemented them
in the \pe{} open-source software framework for massively parallel simulations
of rigid bodies. The \pe{} framework already has been successfully used to simulate
granular systems with and without surrounding fluid in the past~\cite{fischermeier2014simulation, bogner15}.

The discretization of the equations of motion underlying the time-stepping scheme
use an integrator of order one. Contacts are modelled as inelastic and hard
contacts with Coulomb friction. The hard contact model avoids the necessity to
resolve the collision micro-dynamics and the time-stepping scheme avoids the
necessity to resolve impulsive events in time. The one-step integration can be
split into the integration of the velocities and the subsequent
integration of the positions and orientations.

The velocity integration requires
the solution of a non-linear system of equations per time step. In order to
reduce the size of the system in the first place conventional broad-phase
contact detection algorithms are applied to exclude contacts between
intersection hulls. To solve the non-linear system of equations the subdomain
non-linear block Gauss-Seidel is used. The numerical solution algorithm is a
mixture between a non-linear block Gauss-Seidel~(NBGS) and a non-linear block Jacobi with
underrelaxation. In contrast to a pure non-linear block Jacobi it only requires
a mild underrelaxation and in contrast to a non-linear block Gauss-Seidel it
accommodates the subdomain structure of the domain partitioning
and thus allows an efficient parallelization avoiding irregular data
dependencies across subdomains. The implementation of the subdomain NBGS in the \pe{} is matrix-free
and thus avoids the expensive assembly of the Delassus operator. Furthermore, the use of
accumulators and correction variables enables the evaluation of the
particle wrenches in parallel, reuses partial results and reduces the number of
particles that need to be synchronized.

The integration of the positions and orientations is entailed by the execution
of an exceptionally robust synchronization protocol. The key to obtain this
robustness is to add the rank of the parent process and the ranks of the shadow
copy holders to the state of each particle and to explicitly communicate the
state changes. Only then processes can reliably agree upon responsibilities such as
contact treatment and particle integration without being susceptible to numerical errors.

Beyond that, all messages are aggressively aggregated in order to reduce the
communication overhead of small messages and all messages are restricted to
nearest neighbors. The latter is achieved by splitting bodies into local and global bodies and
identifying appropriate requirements. Both measures improve the scalability of
the implementation.

Finally, the scalability was demonstrated for dilute and dense setups on three
clusters, two of them having been in the top~10 of the world's largest
publicly available supercomputers. The parallel efficiency on 
Juqueen
is outstanding. 
T he inter-island scaling results 
on 
SuperMUC 
are 
satisfactory, however, the
intra-island scaling results show room 
for improvement. 
That this is not inherently 
caused by the parallelization
approach can be seen by inspecting the results of the Emmy cluster, whose
architecture is close to a single island of 
SuperMUC.

The largest scaling
experiments
demonstrate that simulations of unprecedented scale with up to $2.8 \cdot 10^{10}$
non-spherical particles and up to $1.1 \cdot 10^{10}$ contacts are possible using up to
$1.8\cdot10^{6}$ processes. The systmatic evaluation also confirms
that 
good parallel efficiency can
be expected on millions of processes even if only a few
hundred particles are allocated to each 
process
provided that the computation exhibits a sufficiently high computational
intensity and the architecture has a good interconnect network .


The favourable scalability results 
do not account for the fact that the NBGS solver does not scale
(algorithmically)
in terms of the number of iterations needed to achieve a given error bound.
Possible future developments arise out of that: The convergence rate of
multigrid methods is independent of the number of unknowns and is in that sense
optimal. A successful construction of such a multigrid method for hard contact
problems would be invaluable for simulating every-increasing system sizes.

%
%
%

\bibliographystyle{spbasic}
\bibliography{2014_Preclik_NSGranularDynamics}

\end{document}